\documentclass[final,5p,times,twocolumn]{elsarticle}

\usepackage{amsmath}
\usepackage{amssymb}
\usepackage{color}   
\usepackage{lipsum}

\journal{Nuclear Physics B}
\usepackage{mdframed}
\usepackage{simplewick}
\usepackage{slashed}
\usepackage{bm}

\usepackage{hyperref}
\hypersetup{hidelinks}
\newcommand{\U}[1]{{\rm U(#1)}}
\newcommand{\SU}[1]{{\rm SU(#1)}}
\newcommand{\SO}[1]{{\rm SO(#1)}}
\newcommand{\eq}[1]{\begin{align}#1\end{align}}
\newcommand{\pmat}[1]{\begin{pmatrix}#1\end{pmatrix}}
\newcommand{\ba}{\begin{array}}
\newcommand{\ea}{\end{array}}
\newcommand{\bma}{\begin{pmatrix}}
\newcommand{\ema}{\end{pmatrix}}
\newcommand{\dis}{\displaystyle}

\def\lag{{\cal L}}
\def\d{\partial}
\def\dd{{\rm d}}

\def\ii{{\rm i}}
\def\e{{\rm e}}
\def\braket#1#2{\left\langle #1\vphantom{#2} \right. \kern-2.5pt\left| #2\vphantom{#1}\right\rangle }
\def\ketbra#1#2{\left| #1\vphantom{#2} \right\rangle \kern-2.5pt\left\langle #2\vphantom{#1}\right| }
\def\bra#1{\left\langle #1\right| }
\def\ket#1{\left| #1\right\rangle }

\def\BraKet#1#2{\left\langle #1\vphantom{#2} \right. \kern-2.5pt\left\| #2\vphantom{#1}\right\rangle }

\newcommand{\nn}{\nonumber}
\newcommand{\lsim}{\raisebox{-0.13cm}{~\shortstack{$<$ \\[-0.07cm] $\sim$}}~}
\newcommand{\gsim}{\raisebox{-0.13cm}{~\shortstack{$>$ \\[-0.07cm] $\sim$}}~}
        \DeclareSymbolFont{CMletters}{OML}{cmm}{m}{it}
        \DeclareMathSymbol{v}{\mathord}{CMletters}{`v}
        \DeclareSymbolFont{epsilon}{OML}{ntxmi}{m}{it}
        \DeclareMathSymbol{\epsilon}{\mathord}{epsilon}{"0F}

\begin{document}

\begin{frontmatter}

\title{\Large Steven Weinberg and Higgs Physics}
\author{{\large\sc Abdelhak Djouadi} and {\large\sc Jos\'e Ignacio Illana}}
\affiliation[first]{
            organization={Departamento de Fisica Teorica y del Cosmos},
            addressline={Universidad de Granada}, 
            postcode={E-18071}, 
            city={Granada},
            country={Spain}           }

\begin{abstract}
As a tribute to  Steven Weinberg, we summarize the immense impact that he had on the understanding of the mechanism of spontaneous symmetry breaking and on the physics of the Higgs boson. In particular, four landmark contributions to this field are highlighted. A first one is his early work with Goldstone and Salam on spontaneously broken continuous symmetries that paved the way to the Higgs mechanism. A second towering breakthrough is his model of leptons which later became the Standard Model of particle physics and for which he was awarded the Nobel prize with Glashow and Salam. A third seminal work is the so-called Weinberg-Linde lower bound on the Higgs boson mass that was derived from the requirement of the stability of the electroweak vacuum. Finally, we summarize his important contributions in model-building of new physics with extended Higgs sectors and their possible impact in flavor physics and CP-violation. The historical aspects as well as the contemporary way of viewing these four major topics are summarized and their impact on today Higgs physics, and more generally particle  physics, is highlighted.    
\end{abstract}

\end{frontmatter}

\section{Introduction}
\vspace*{-1mm}
\label{introduction}

Steven Weinberg was a giant of our field. Without a single doubt, he was one of the leading  and most influential figures in particle physics during the second half of the twentieth century. He played a key role in the making of the Standard Model (SM) of elementary particles, the theory that describes three of the four forces in Nature, the strong, the weak and the electromagnetic interactions. He also made a major enlightenment in the description and the understanding of the fourth force, the gravitational one. Despite of the fact that the SM, together with the theory of general relativity for gravitation,  explains most if not all phenomena observed at currently probed energy scales, he nevertheless believed that the model ``has too many arbitrary features for its predictions to be taken very seriously". 

He eventually embarked in the construction and development of new physics theories that go beyond the SM. He pioneered or made landmark achievements in the most important directions, ranging from Grand Unified Theories, to supersymmetric models including supergravity, to strongly interacting theories with dynamical symmetry breaking and, even, to models with extra space-time dimensions. He was also  instrumental in promoting the alternative proposal of ``effective field theories" to describe and test new physics effects at low energies and first applied it to explain the mass of the neutrinos and their smallness, through what is now called the seesaw mechanism. This effective approach is used intensively in present days and is considered by many to be the central tool to test the triumphant SM and to indirectly search for tiny new physics effects beyond it.

Furthermore, Weinberg left a determining mark in the field of cosmology and in astroparticle physics. He made transparent the point that particle physics plays a crucial role in the study of the early universe, soon after the Big Bang. He introduced and promoted several fundamental ideas and subjects, such as the role of Grand Unification in the possibility of proton decay and in the observed baryon asymmetry in the Universe, and studied the cosmological implications of non-zero or finite-temperature effects in renormalizable quantum field theories. He also worked on the extremely arduous problem of the cosmological constant, proposed an ``anthropic" solution to explain its smallness and clarified the role of the massless spin-2 graviton in the mediation of the gravitational interaction. 

Finally, besides being an immense physicist, Weinberg was a great pedagogue, a preeminent intellectual and, occasionally, a public spokesperson for physics and science in general. He shaped and clarified the perceptions of a generation of students and researchers by his textbooks such as the seminal ``{Gravitation and Cosmology: principles and applications of the general theory of relativity}" and even of the general public by his science books such as the very popular and best-seller ``{The First Three Minutes: a modern view of the origin of the Universe}" and ``{Dreams of a Final Theory: the scientist's search for the ultimate laws of Nature }" which, although remaining faithful to physics, were accessible to alert non-scientist readers. More recently, in 2015, he even extended his thinking to the history of sciences and wrote the daring and illuminating ``{To Explain the World: The Discovery of Modern Science}" which initiated many debates and intensive discussions.   

In this homage to the memory of Steven Weinberg who passed away in July 2021, we focus on the pioneering and groundbreaking contributions that he made from the early sixties to the late seventies in the area of spontaneous symmetry breaking and the physics of the Higgs boson and briefly summarize them. The Higgs mechanism, together with the principle of gauge symmetry, is one of the cornerstones of the SM which has been triumphantly confirmed by thundering experiments, including the discovery of the Higgs boson a decade ago.  

A first seminal contribution of Weinberg that we summarize is his pioneering work on the mechanism of spontaneous breaking in theories invariant under continuous global symmetries \cite{Goldstone:1962es}. In this mechanism, while the Lagrangian is invariant under the global symmetry, the vacuum or ground state is not; the symmetry is  then said to be hidden or spontaneously broken. In a key paper written with Jeffrey Goldstone and Abdus Salam, Weinberg made a systematical investigation of the phenomenon and proved a conjecture, which became later known as the Goldstone theorem, which states that there is a massless mode, a spinless particle called a Goldstone boson, for each broken generator of the symmetry, i.e.  that does not preserve the ground state. This theorem had an immense impact and, circumventing it, was the main motivation to promote the initial global symmetry to a local gauge symmetry, which led to the Higgs mechanism that generates masses for gauge bosons.  

The second of Weinberg's contributions that we outline is his 1967 dramatic and historical formulation of the unified theory of the weak and electromagnetic interactions \cite{Weinberg:1967tq}. This ``model of leptons" (as there was not yet an established theory for quarks), witnessed the fortunate marriage of the ${\SU2_L \times \U1_Y}$ gauge symmetry for the electroweak interaction, following the Yang-Mills idea  of a non-abelian \SU2 gauge theory for isospin conservation in strong interactions,  and the mechanism of spontaneous symmetry breaking or Higgs mechanism.  The latter allowed to gives masses in a gauge invariant manner not only to the weak $W^\pm$ and newly posited $Z^0$ gauge bosons, the mediators of the short range weak force, but also to the leptons and quarks. Hence, the modern concept of the ``Higgs mechanism" to generate particle masses is also due to Weinberg as he was the first to apply it to fermions. This is the work that earned him in 1979 the Nobel Prize in Physics, shared with Sheldon Glashow and Abdus Salam.

A third major contribution made by Weinberg in 1975 concerns the mass of the Higgs boson, the spin-zero relic of his electroweak model. At a time where little attention was given to this particle and where there was essentially no constrain on its properties, he proposed the so-called vacuum stability lower bound on the Higgs mass \cite{Weinberg:1976pe} from the constraint that, when including quantum corrections,  the scalar potential that breaks the ${\SU2_L \times \U1_Y}$ symmetry should not develop a minimum that is deeper than the electroweak one. This Linde-Weinberg lower bound, together with the upper bound of about 1~TeV from constraints from perturbativity and unitarity, allowed to corner the Higgs boson in a narrow mass range and greatly facilitated the studies that allowed for its discovery at the LHC.  

Finally, we will briefly summarize some of Weinberg's work in extensions of the Higgs sector to more than the single doublet that is needed in the SM. He played a pioneering role in this context since, as early as 1976, he proposed with Gildener a two-Higgs doublet model (2HDM) that provides naturally a light and SM-like (or aligned) Higgs boson \cite{Gildener:1976ih}. A year later, in a seminal paper with Sheldon Glashow \cite{Glashow:1976nt}, he introduced discrete symmetries that provided major constraints on how the known fermions should couple to Higgs bosons in multi-Higgs doublet models without generating unwanted flavor changing neutral currents. He also  contemplated the possibility that CP-violation arises purely from Higgs boson exchange \cite{Weinberg:1976hu} and studied its consequences; see Refs.~\cite{Weinberg:1989dx,Weinberg:1990me}.
 
In a concluding  section, we summarize the following developments of all these ideas in the context of Higgs and SM  physics, which were crowned by the spectacular discovery of the Higgs particle  \cite{Aad:2012tfa,Chatrchyan:2012xdj} at the CERN LHC in July 2012 and earned the Nobel prize to Englert and Higgs a year later. Thereafter, the profile of the particle was  determined  and its couplings precisely measured \cite{ATLAS:2022vkf,CMS:2022dwd}. These results led to the inference that the SM, half of a century after its birth,  is indeed the correct theory that allows to describe all phenomena at energies below the TeV scale \cite{Weinberg:2018apv} and that, unfortunately, new physics beyond it, including models that Weinberg himself introduced and studied (such as theories with strong dynamical symmetry breaking  \cite{Weinberg:1975gm} and supersymmetric models \cite{Weinberg:1981wj,Farrar:1982te,Hall:1983iz}), is severely constrained by data and should eventually manifest itself  only at much higher energy scales. 

\section{Broken Symmetries}

In the early 1960's, many theorists including Weinberg were trying to make sense of the confusing field of strong interactions with its many observed hadronic resonances. A challenge was to understand the origin and implications of approximate sym\-metries that were apparently behind it, such as the Gell-Mann's ``eightfold way" \cite{Gell-Mann:1961omu} that led to a successful classification of the lightest hadrons. The idea of spontaneous symmetry breaking  was a very appealing possibility to somewhat clarify the issue. In this mechanism, imported in the early sixties by Yoichiro Nambu \cite{Nambu:1960tm} from  the theory of superconductivity in condensed matter physics and concretely applied to particle physics by Jeffrey Goldstone \cite{Goldstone:1961eq} shortly after, there is a global symmetry under which the fields and their interactions transform in an invariant way but the vacuum is not symmetric. The original symmetry is not apparent anymore and is dubbed spontaneously broken. However, already at a rather early stage, it gave rise to a severe problem as it implied the existence of massless scalar particles that have not been experimentally observed. 

\begin{figure}[!ht]
\centering
\vspace*{1mm}
\begin{tabular}{cc}
\includegraphics[scale=0.33]{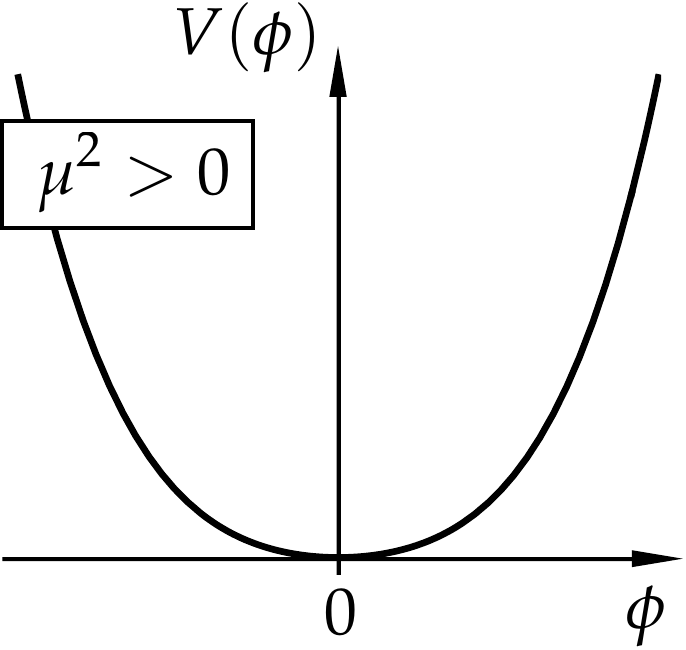} &\quad
\includegraphics[scale=0.33]{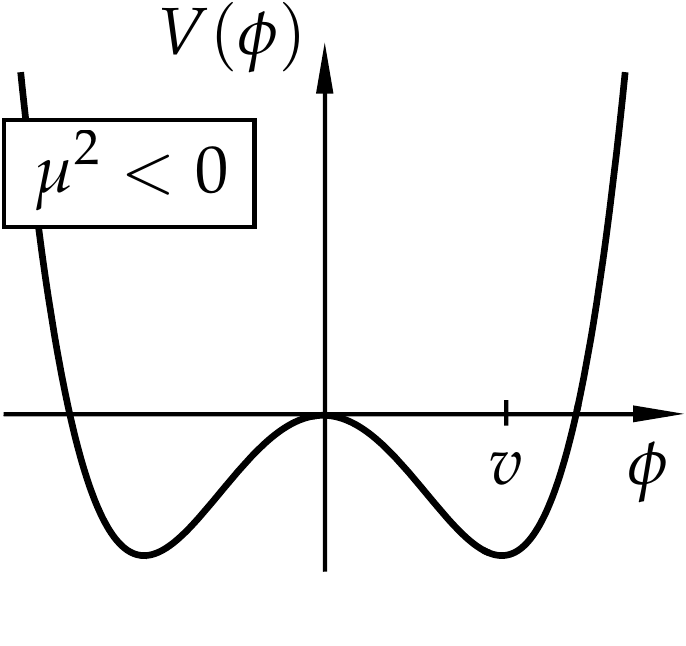} 
\end{tabular}
\vspace*{-1mm}
	\caption{The potential $V(\phi)$ of a real scalar field, symmetric under the transformation $\phi \mapsto -\phi$, for positive (left) and negative (right) mass squared term.}
	\label{fig:SSB}
\vspace*{-1mm}
\end{figure}

To understand the concept of spontaneous symmetry breaking (SSB),  consider a real scalar field $\phi$ with a Lagrangian
\begin{equation}
\lag = \frac{1}{2}(\d_\mu\phi)(\d^\mu\phi) - V(\phi)\,,\quad
V(\phi) = \frac{1}{2}\mu^2\phi^2 + \frac{\lambda}{4}\phi^4,
\label{eq:Vreal}
\end{equation}
invariant under a discrete $\mathbb{Z}_2$ symmetry $\phi \mapsto -\phi$. $\lag$ is hermitian if the parameters $\mu^2$ and $\lambda$ are real and $\lambda>0$ ensures there exists a ground state. Two cases can be distinguished depending on the sign of $\mu^2$; see Fig.~\ref{fig:SSB}. For $\mu^2 >0$, the field $\phi$ has a minimum at $\phi\!=\!0$ and simply corresponds to a scalar boson with mass $\mu$.

More interesting  is the case with $\mu^2\!<\!0$ for which the mi\-nimum of the field $\phi$ is degenerate and non-zero, $\phi \! = \! v \! \equiv \! \pm\sqrt{-\mu^2/\lambda}$. For a quantum field, the configuration of minimum   energy is interpreted as its vacuum expectation value (vev) $\bra{0}\phi\ket{0}\!=\!v$.  The physical quantum field denoted $\eta$ with $\bra{0}\eta\ket{0} = 0$ is then obtained by the field redefinition $\phi \! \equiv \! v \! + \! \eta$; it describes excitations over the vacuum and has a Lagrangian 
\begin{align}
\lag &= \frac{1}{2}(\d_\mu\eta)(\d^\mu\eta)- \lambda v^2 \eta^2  -\lambda v \eta^3 -\frac{\lambda}{4}\eta^4\,,
\label{lag:discrete}
\end{align}
and a mass $m_\eta=\sqrt{2\lambda v^2}$. As can be seen from eq.~(\ref{lag:discrete}), the $\mathbb{Z}_2$ symmetry of the original Lagrangian is broken or,  to be more precise, hidden. Indeed, the Lagrangian above does not exhibit an explicit symmetry breaking since the coefficients of the terms $\eta^2$, $\eta^3$ and $\eta^4$ are not independent and are determined by the two parameters $\lambda$ and $v$, a remnant of the original symmetry. We say that the symmetry is spontaneously broken because it is due to a non-invariant vacuum, not to an external agent. 

\begin{figure}[!ht]
\centering
\vspace*{1mm}
\includegraphics[scale=0.4]{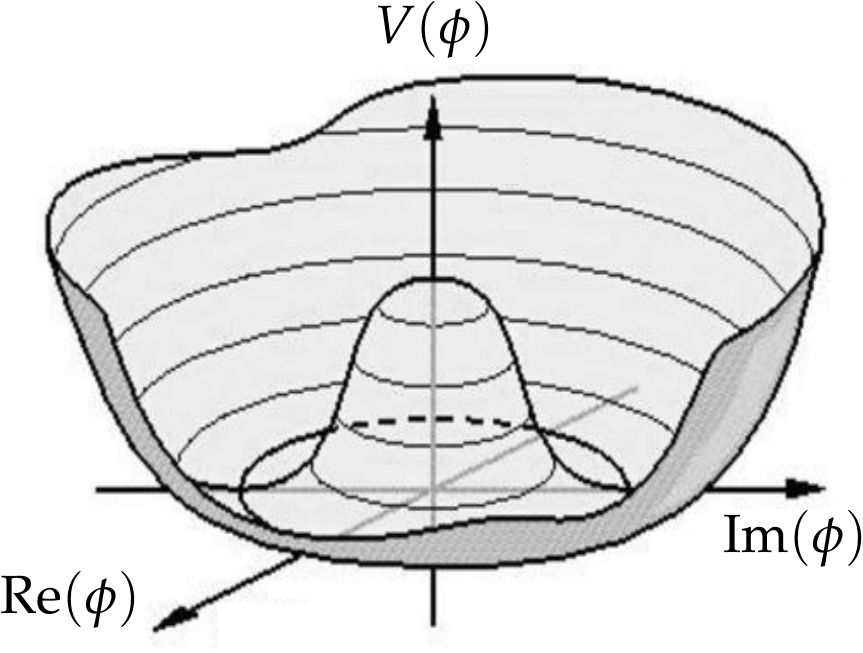}
\vspace*{-1mm}
 \caption{The potential $V(\phi)$ as a function of the real and imaginary components of the complex field $\phi$
 exhibiting a non-zero vacuum expectation value.} 
	\label{fig:Mexican}
\vspace*{-1mm}
\end{figure}

The Goldstone conjecture,  the appearance of massless  scalar particles,  which was promoted to a theorem in Ref.~\cite{Goldstone:1962es}, comes when one considers a global rather than a discrete symmetry to which obeys a complex scalar field $\phi$ with a Lagrangian
\begin{equation}
\lag = (\d_\mu\phi)^\dagger(\d^\mu\phi) - V(\phi)\,,\quad
V(\phi) = \mu^2\phi^\dagger\phi + \lambda(\phi^\dagger\phi)^2\,,
\label{Lag:continuous}
\end{equation}
which is invariant under the global $\U1$ transformations $\phi \mapsto \e^{-\ii Q\theta}\phi$. In the $\lambda \! > \!0$, $\mu^2\! <\!0$ configuration,  the potential has the Mexican hat shape shown in Fig.~\ref{fig:Mexican} with a degenerate minimum, $\bra{0}\phi\ket{0} \equiv v/\sqrt{2}$ with the vev $v=\sqrt{-\mu^2/\lambda}$ taken to be real and positive without loss of generality.   Upon the field redefinition   
\begin{align}
\phi\equiv  [v+\eta+\ii\chi ]/\! \sqrt 2 , \quad \mbox{with} ~ \bra{0}\eta\ket{0}=\bra{0}\chi\ket{0}=0\,,
\end{align}
the Lagrangian is no longer invariant under the $\U1$ symmetry,
\begin{equation}
\lag = \frac{1}{2}(\d_\mu\eta\d^\mu\eta) 
      \!+\!\frac{1}{2}(\d_\mu\chi\d^\mu\chi)
      \!-\!\lambda v^2\eta^2
      \!-\!\lambda v \eta(\eta^2+\chi^2)
      \!-\!\frac{\lambda}{4}(\eta^2\!+\!\chi^2)^2.
\label{Lag:physical}
\end{equation}
The spontaneous breaking of this continuous symmetry leaves one massless scalar field $\chi$,  whereas $\eta$ is massive, $m_\eta=\sqrt{2\lambda}v$.

The consequences of the spontaneous breaking above are more evident in the case of a group with more symmetries. For instance, take a triplet of real scalar fields $\Phi$ whose self-interactions are again given by a Mexican hat potential 
\begin{equation}
\lag = \frac{1}{2}(\d_\mu\Phi)^{T}(\d^\mu\Phi) -\frac{1}{2}\mu^2\Phi^{T}\Phi - \frac{\lambda}{4}(\Phi^{T}\Phi)^2,
\end{equation}
that is invariant under global $\SO3$ transformations $\Phi \mapsto \e^{-\ii T_a\theta^a}\Phi$ where $T_a$ are the three generators of the group.\footnote{The number of generators is $N$ for the SO($N$) group and $N^2-1$ for the SU($N$) group. Hence, SO(3) and \SU2 have 3 generators which correspond to a half of the the Pauli matrices, $T_a = \frac12 \tau_a$ with  $a={1,2,3}$ in the doublet representation}. Again, for $\lambda\!>\!0$ and $\mu^2\!<\!0$, one has $\bra{0}\Phi^{T}\Phi\ket{0} = v^2 = -\mu^2/\lambda$ and may express the field as $\Phi \equiv \big(\varphi_1,\, \varphi_2,\, v + \varphi_3\big)^{T}$. Also defining the complex combination $\varphi \equiv (\varphi_1+\ii\varphi_2)/\sqrt{2}$, one then has 
\begin{align}
\lag &= (\d_\mu\varphi)^\dagger(\d^\mu\varphi) +\frac{1}{2}(\d_\mu\varphi_3)(\d^\mu\varphi_3)
     -\lambda v^2\varphi_3^2\nonumber\\
    & \quad-\lambda v (2\varphi^\dagger\varphi + \varphi_3^2)\varphi_3 -\frac{\lambda}{4}(2\varphi^\dagger\varphi + \varphi_3^2)^2 \,.
\end{align}
The Lagrangian is no longer symmetric under $\SO3$ but is invariant under the $\U1$ transformation $\varphi \mapsto \e^{-\ii Q\theta}\varphi$ with $Q$ arbitrary and  $\varphi_3 \mapsto \varphi_3$ ($Q=0$ if viewed as above). 

In other words, the group $\SO3$ has broken spontaneously into a $U1$ subgroup. Since there are $3-1=2$ broken generators, two real scalar fields (or, equivalently, one complex scalar $\varphi$) remain massless, while the other scalar, the field  $\varphi_3$,  acquires a mass proportional to $v$, $m_{\varphi_3}=\sqrt{2\lambda v^2}$.

The examples above illustrate the Goldstone conjecture: the number of massless particles or Goldstone bosons equals the number of spontaneously broken generators of the symmetry. 

In their seminal paper \cite{Goldstone:1962es}, Goldstone, Salam and Weinberg presented not just one but three proofs of this conjecture which then became a theorem. Rather than giving the proofs here, let us provide,  instead, a simple explanation. By definition of a symmetry, if the Hamiltonian ${\cal H}$ is invariant under a Lie group, it commutes with the generators $T_a$ of the group $[T_a,{\cal H} ] = 0$ with  $a=1,\dots, N$.  Also, by definition of the vacuum state, one has ${\cal H} \ket{0} = 0$ and thus, ${\cal H} (T_a\ket{0})=T_a {\cal H} \ket{0}=0$. As a consequence,  if $\ket{0}$ is such that $T_a\ket{0}=0$ for all generators, there is a single minimum: the vacuum, that will remain invariant.  But if the vacuum $\ket{0}$ is such that $T_{a'}\ket{0}\ne 0$ for some (broken) generators $T_{a'}$, the minimum is degenerate: for any choice (true vacuum) we will have $\e^{-\ii T_{a'}\theta^{a'}}\ket{0} \ne \ket{0}$, so it will not remain invariant. In this case, there are excitations from $\ket{0}$ to $\e^{-\ii T_{a'}\theta^{a'}}\ket{0}$ (flat directions of the potential) that cost no energy, hence corresponding to massless particles, the Goldstones. 

The proof of the Goldstone theorem in 1962, which implied the existence of massless bosons rather than the desired massive ones to be identified with hadrons, was a severe setback. Quoting Shakespeare's King Lear, Weinberg wrote ``Nothing will come of nothing". Nevertheless, only two years afterwards, it was found that there is a simple way to evade the theorem: making the symmetry local, as it is the case of electromagnetic gauge invariance. In this case, the would be Goldstone bosons produced by SSB  can be eliminated by a gauge transformation. 

To see how the SSB mechanism works in this case, take the gauge invariant Lagrangian for a complex scalar field $\phi$ 
\eq{
\lag =  -\frac{1}{4}F_{\mu\nu}F^{\mu\nu} + (D_\mu\phi)^\dagger(D^\mu\phi)
      -\mu^2\phi^\dagger\phi - \lambda(\phi^\dagger\phi)^2 \, , 
}
with $F_{\mu\nu}$ the strength field tensor and  $ D_\mu \! \equiv\! \d_\mu \!+\! \ii e Q A_\mu$ the covariant derivative, that is invariant under local \U1 transformations
\begin{eqnarray}
\phi(x) \mapsto \e^{-\ii Q\theta(x)}\phi(x) \,, \quad
A_\mu(x) \mapsto A_\mu(x) + \dis {e^{-1}}\d_\mu\theta(x).
\end{eqnarray}
Again, if $\lambda\!>\!0$ and $\mu^2\!<\!0$, the potential has the Mexican hat shape of Fig.~\ref{fig:Mexican} with a minimum at $\bra{0}\phi \ket{0}=v/\sqrt{2}$ where $v=\sqrt{-\mu^2/\lambda}>0$. As done before, we rewrite the field $\phi$ in terms of  two real fields $\eta$ and $\chi$ with null vevs, choosing this time the parametrization $\phi \equiv \e^{\ii \chi/v}[v+\eta]/\sqrt 2$, so that the field $\chi$ can be eliminated (or gauged away) by exploiting the gauge freedom:  
\begin{align}
\phi(x) \mapsto \e^{-\ii \chi(x)/v}\phi(x) =  [v+\eta(x)]/\! \sqrt2.
\end{align}
In this (unitary) gauge, the previous Lagrangian becomes
\eq{
\lag =& -\tfrac{1}{4}F_{\mu\nu}F^{\mu\nu} +\tfrac{1}{2}(eQv)^2 A_\mu A^\mu
        +\tfrac{1}{2}(\d_\mu\eta)(\d^\mu\eta) -\lambda v^2\eta^2 \nn\\
     & -\lambda v\eta^3-\tfrac{\lambda}{4}\eta^4
       +\tfrac{1}{2}(eQ)^2 A_\mu A^\mu(2v\eta+\eta^2)\,.
}
As can be seen, the gauge boson $A_\mu$ has acquired a mass $M_A = |eQv|$. The scalar field $\eta$ has also a mass $m_{\eta}=\sqrt{2\lambda}\, v$, so there is no Goldstone boson. In fact,  the photon (with two degrees of freedom) has absorbed the would-be Goldstone boson (with one degree of freedom) and became massive (i.e. with three degrees of freedom): the additional longitudinal polarization is the Goldstone boson. The \U1 gauge symmetry is no more apparent and we again say that it is spontaneously broken.  

The results above are the simplest application of the Brout-Englert-Higgs mechanism 
\cite{Higgs:1964ia,Englert:1964et,Guralnik:1964eu,Higgs:1964pj,Higgs:1966ev} (or Higgs mechanism for short): the gauge bosons associated with the spontaneously broken generators become massive, the corresponding would-be Goldstone bosons (one per broken symmetry) are unphysical (they can be absorbed), and the remaining massive scalars (a single Higgs boson in this case) are physical. The existence of a Higgs boson is the smoking gun confirming that this mechanism is responsible for the mass of the gauge bosons associated to broken symmetries.  The Goldstone field is eliminated but the number of degrees of freedom of the physical spectrum (four  in the  case discussed here) remains the same. 

A few years after the Higgs mechanism was introduced, Weinberg tried to apply it to the strong interactions, imagining that chiral symmetry was perhaps not global but local and the interaction could be described by a spontaneously broken  ${\SU2_{L}\times \SU2_{R}}$ gauge theory broken to the vector symmetry ${\SU2_{V}}$. The $\rho$ and $a_1$ mesons were then the massless and massive gauge bosons, respectively. However, an explicit mass term had to be added by hand to reproduce the $\rho$ meson mass,  hence breaking explicitly gauge invariance and spoiling the conjectured renormalizability of the theory. But soon afterwards, he realized that he was applying the right idea to the wrong problem: it was the weak interactions instead that could be described by a spontaneously broken gauge theory choosing the appropriate symmetry group, and it was not the $a_1$ mesons but the weak bosons that acquire a mass. This led to the electroweak standard model to which we turn now. 

\section{The Model for Leptons}

From the beginning, the weak interaction had a special status as it was of short range nature, besides the fact that it was violating parity. The old theory of Fermi with the four-fermion contact interaction \cite{Fermi:1934hr} describing the $\beta$ decay, required more than three decades  to be turned into a universal interaction mediated by the massive $W^\pm$ vector bosons \cite{Feynman:1958ty,Sudarshan:1958vf}, just in the same way as the electromagnetic interaction proceeds via the exchange of massless photons $\gamma$. The proposed vector-axialvector structure of the weak interaction resolved the effective Fermi constant $G_F$ in terms of a fundamental weak coupling $g$, analogous to the electromagnetic one $ e$, and the $W^\pm$ boson mass. 

Because of this mass put by hand, the new theory was nevertheless not renormalizable and even not unitary at energies much above the Fermi scale of a few hundred GeV. The aspiration to describe the weak  interaction in terms of a gauge theory with a symmetry group larger than the \U1 electromagnetic group, similar to the non-abelian generalized isospin \SU2 symmetry group that Yang and Mills introduced in 1954 \cite{Yang:1954ek} to describe the strong interactions, hence collapsed because the $W^\pm$ mediators were massive in contrast to the vector bosons of gauge theories like the photon which should be massless.\footnote{It is worth remembering that the roots of gauge invariance \cite{Jackson:2001ia} go back to early 19th century, even before Maxwell formulated his theory of the electromagnetic field. However, it was Weyl who introduced the word ‘gauge’ in the 1920s and consecrated
the gauge symmetry as the guiding principle for the construction of interacting quantum field theories, enshrining the modern gauge principle in which the gauge four-vector fields follow from the requirement of gauge invariance of matter field equations under gauge transformations.}  

This is what Sheldon Glashow attempted as early as  1961 \cite{Glashow:1961tr} when he proposed a model based  on an ${\SU2\times \U1}$ gauge symmetry.\footnote{Salam and Ward \cite{Salam:1964ry} also proposed a rather similar model in 1964. In fact, Schwinger had a little earlier proposed a first unified gauge theory of weak and electromagnetic interactions \cite{Schwinger:1957em} involving both the $W^\pm$ and $\gamma$ states.} This model was already unifying the electromagnetic and weak interactions and, besides the $W^\pm$ and the photon, a fourth gauge boson, the $Z^0$ coupled to a weak neutral current, had to be introduced to complete the set of states that correspond to the four generators of the symmetry group. The photon and $Z^0$ states resulted from the mixing of the two neutral gauge bosons with angle\footnote{Ironically, the $\theta_W$ introduced by Glashow is often called Weinberg angle.} $\theta_W$, and the $W^\pm$ bosons mediated the weak charged current interaction with coupling $g=e/\sin\theta_W$. Of course, this first unification failed simply because one has to introduce masses for $W^\pm$ and $Z^0$ by hand to make the weak interaction of short range, spoiling the ${\SU2\times \U1}$ gauge invariance and, hence, the nice properties of the theory such as renormalizability. The Glashow model did also not address the problem of the non-zero fermion masses. 

The gauge boson and the fermion mass problems were solved by Weinberg in 1967  \cite{Weinberg:1967tq} (and independently by Salam  who published his work the following year \cite{Salam:1968rm}) by using for the first time the Higgs mechanism to spontaneously break the ${\SU2\times \U1}$ gauge group of the unified electroweak interactions. In fact, in his paper, Weinberg concedes that his model is similar to the one of Glashow and ``the chief difference is that Glashow introduces symmetry-breaking terms into the Lagrangian, and therefore gets less definite predictions''. Let us describe the main aspects of Weinberg's work. (For a detailed discussion on the structure of the SM see, e.g., Ref.~\cite{Illana:2022hab}.)

The electromagnetic and weak interactions combine into the electroweak interaction based on the symmetry group ${ \SU2_{L} \times \U1_{Y}}$. The fermions of each family appear in two configurations: left-handed fermions $f_L$ in doublets of weak isospin and right-handed fermions $f_R$ in iso-singlets. Because at that time there was no theory for quarks (their existence became questionless only a few years later), Weinberg specialized to the case of first generation leptons: the left-handed electron and its neutrino are in a doublet $L_L=(\nu_L\; e_L)^T$
of weak isospin third components $I_3=\pm\frac12$, respectively, while the right-handed electron $e_R$ appears as a singlet of $I_3=0$; there is no right-handed neutrino. The quantum number of hypercharge $Y_f$ is given  by the  electric charge and the weak isospin, $Y_f=Q_f-I_3^f$.  

The interaction is mediated by the exchange of four gauge  bosons:  $W^\mu_{1,2,3}$ corresponding to the 3 generators of ${\SU2_{L}}$, to be identified with the three $2\!\times \!2$ Pauli matrices $\tau_{1,2,3}$,   and $B_\mu$ corresponding to the generator $Y$ of ${\U1_Y}$. The physical gauge bosons, $W^\pm$, $Z^0$ and $\gamma$, are linear combinations of these: 
\eq{
W^\pm_\mu \!=\!\frac{W^1_\mu \mp \ii W^2_\mu}{\sqrt{2}},\quad 
Z_\mu \!=\! \frac{g W^3_\mu {+} g' B_\mu}{\sqrt{g^2+g'^2}},\quad
A_\mu \!=\! \frac{g' W^3_\mu {-} g B_\mu}{\sqrt{g^2+g'^2}}, 
\label{eq:th:new-fields}
}
with $g$ and $g'$ the couplings of ${\SU2_{ L}}$ and  ${\U1_{Y}}$ that obey
\begin{eqnarray}
e = g \sin\theta_W = g' \cos\theta_W,  \quad
e=gg'/\sqrt{g^2+g'^2}\, . 
\end{eqnarray}
This is the electroweak unification which relates the couplings $g$ of the weak and  $e$ of the electromagnetic interactions through the Weinberg or weak mixing angle $\theta_W$  \cite{Glashow:1961tr}. 

In order to provide the $W^\pm$ and $Z^0$ with masses, without spoiling gauge invariance, Weinberg implemented the Higgs mechanism in such a way that ${\SU2_{L} \times \U1_{Y} \to \U1_{Q}}$, i.e. the symmetries associated to three out of the four generators of the gauge group get spontaneously broken while the one associated to the combination $Q=T_3+Y$ is unbroken, so the photon remains massless. This electroweak symmetry breaking cannot be achieved by just introducing one complex scalar field. Weinberg, and apparently also Nature, chose a complex Higgs doublet of $SU(2)$ with the appropriate hypercharge,
\begin{eqnarray}
\Phi = \pmat{\phi^+ \\ \phi^0}\ , \quad  \bra{0}\Phi\ket{0} \equiv \frac{1}{\sqrt{2}}\pmat{0 \\ v}, \quad
\end{eqnarray}
where the Higgs vev $v/\sqrt{2}$ is the minimum of the potential 
\begin{eqnarray}
V(\Phi) = \mu^2\Phi^\dagger\Phi + \lambda(\Phi^\dagger\Phi)^2\, , 
\label{eq:Vscal}
\end{eqnarray}
with $\mu^2=-\lambda v^2<0$ and $\lambda>0$. The covariant derivative $D_\mu\Phi = (\d_\mu - \ii g \widetilde W_\mu + \ii g' y_\Phi B_\mu)\Phi$ ensures that the Higgs Lagrangian is gauge invariant, introducing interactions with the gauge fields,
\begin{eqnarray}
{\cal L}_\Phi = (D_\mu\Phi)^\dagger D^\mu\Phi - V(\Phi).
\end{eqnarray}
By assigning a hypercharge $y_\Phi=\tfrac{1}{2}$ to the Higgs doublet, the generator associated to the photon field will annihilate the vacuum, in contrast to the other ones, as desired. Consequently, the Lagrangian $\lag_\Phi$ contains weak boson mass terms with
\begin{eqnarray}
M_W=M_Z\cos\theta_W=\tfrac{1}{2}gv.
\end{eqnarray}
This allowed Weinberg to derive the effective Fermi coupling (determined e.g.~from muon lifetime) from the $W$-mediated weak interactions,  $G_F/\sqrt{2}\!=\!g^2/8M_W^2\!=\! 1/2v^2$, predicting $v=(\sqrt{2} G_F)^{-1/2}\approx 246$~GeV, $M_W=e v/2\sin\theta_W\gsim 40$~GeV and $M_Z= e v/\sin2\theta_W\gsim 80$~GeV.
On the other hand, in the unitary gauge, three would-be-Goldstone fields in $\Phi$ get gauged away to become the helicity-zero states of the $W^\pm$ and $Z^0$ bosons. 

The remaining degree out of the four initial degrees of freedom  of the field $\Phi$, corresponds to the physical Higgs field $H$ 
\eq{
  \Phi(x)=\frac{1}{\sqrt{2}}\pmat{0\\ \phi(x)},\quad \phi(x)=v+H(x),
\label{eq:Phiphi}
}
whose mass $M_H=\sqrt{-2\mu^2}=\sqrt{2\lambda}v$ was also given in the original Weinberg's paper, with $\lambda$ being unknown at the time. 

To generate the fermion masses, in the same work, Weinberg pioneered the introduction of a gauge invariant Yukawa interaction of the lepton doublet and singlet with the Higgs doublet,
\eq{
{\cal L}_{\rm Yuk}  &= {-} \lambda_{e} \overline{L}_L \Phi e_R + {\mbox{h.c.}}\,  \nn\\
& \supset {-} \frac{\lambda_e}{\sqrt{2}} \pmat{\overline{\nu}_e & \!\!\overline{e}_L}
                              \pmat{0 \\ v+H}\, e_R
      = {-} \frac{\lambda_e}{\sqrt{2}} (v+H)\, \overline{e}_L e_R\, , 
\label{eq:Yukawa-e}
}
where the Yukawa coupling $\lambda_e$ is not arbitrary but related to the electron mass $m_e= \lambda_e v/\sqrt{2}$. The neutrino remains massless. 

This completes Weinberg's model for leptons in which the weak and the electromagnetic interactions were unified under an \SU2x\U1 gauge symmetry, and both the weak vector boson masses and the fermion masses were generated in  gauge invariant way using the Higgs mechanism. This model became the SM when two major ingredients were later added. 

A first one was the incorporation of quarks. In 1967 the light $u,d,s$ quarks proposed by Gell-Mann and Zweig \cite{Gell-Mann:1961omu,Zweig:1964ruk} were only hypothetical and they became physical entities at SLAC only two years later \cite{Bloom:1969kc,Breidenbach:1969kd}.  Subsequently, the charm-quark, as a partner of the strange-quark, was twice observed \cite{E598:1974sol,SLAC-SP-017:1974ind}. To that, one has to add a third generation of fermions: the tau lepton \cite{Perl:1975bf} was discovered in 1975, the bottom quark \cite{E288:1977xhf} in 1977 and at last in 1995, the very heavy top quark \cite{CDF:1995wbb,D0:1995jca}. 

For any fermion generation, the masses are introduced using the ${\SU2_{L} \times \U1_{Y}}$ invariant Yukawa Lagrangian
\begin{eqnarray}
{\cal L}_{\rm Yuk}  = -\overline{L}_L \lambda_\ell \Phi \ell_R
                     -\overline{Q}_L \lambda_d \Phi d_R
                     -\overline{Q}_L \lambda_u \widetilde\Phi u_R + {\rm h.c.},
\end{eqnarray}
which is a generalization of the Weinberg interaction eq.~(\ref{eq:Yukawa-e}). The conjugate field $\widetilde\Phi\equiv i\sigma_2\Phi^*$ has the appropriate quantum numbers for interactions involving up-type fermion singlets. After symmetry breaking, the fermion masses are proportional to the corresponding Yukawa couplings, $m_f=\lambda_f v/\sqrt{2}$.  For three fermions generations, these couplings $\lambda_f$ are actually matrices in flavor space, that upon diagonalization yield mass eigenstates and, in the quark sector, enable quark mixing that shows up in charged-current interactions mediated by the $W^\pm$. For three generations of quarks, this Cabibbo-Kobayashi-Maskawa (CKM) quark mixing matrix \cite{Cabibbo:1963yz,Kobayashi:1973fv} has a complex phase, which is the only source of CP violation of the SM. But there are no flavor changing neutral currents at tree level, that appear loop-suppressed by the Glashow-Iliopoulos-Maiani (GIM) mechanism  \cite{Glashow:1970gm} that predicted the charm quark. 

It is worth noting that the Higgs coupling to gauge bosons is proportional to the masses squared $\lag_\Phi \supset M_W^2 W_\mu^\dagger W^\mu \left(1+{H}/{v}\right)^2$ and the coupling to fermions is proportional to the masses, ${\cal L}_{\rm Yuk} \supset -m_f\bar{f} f \left(1+{H}/{v}\right)$. This is a distinguishing feature of the Higgs boson that constitutes its fingerprint.

The other important issue, prior to its acceptance,  was the renormalizability of the theory. Yang-Mills theories had been proven renormalizable \cite{Faddeev:1967fc} and Weinberg (as well as Salam) was convinced  that this property was kept by spontaneous breaking but he did not find a way to prove it. According to him, the problem was that he had adopted the ``too obscure'' unitary gauge. The proof was provided in 1971 by 't Hooft and Veltman \cite{tHooft:1971qjg,tHooft:1972tcz} (and later by Lee and Zinn-Justin \cite{Lee:1972ocr}), who introduced the more practical Feynman-'t Hooft gauge. The model of Glashow, Weinberg and Salam was then predictable\footnote{According to Sidney Coleman, ``the proof of renormalizability of 't Hooft turned the Weinberg-Salam frog into an enchanted prince" \cite{Coleman:1979}.} and became the SM of the electroweak interactions; see e.g.~\cite{Abers:1973qs}. They were awarded the 1979 Nobel Prize in Physics ``for their contributions to the unification of the weak and electromagnetic interaction between elementary particles''. This happened after the discovery of weak neutral currents in neutrino scattering by the Gargamelle collaboration in 1973 \cite{GargamelleNeutrino:1973jyy}, but even before the $W^\pm$ and $Z^0$ gauge bosons were discovered in the CERN Super Proton Synchrotron collider in 1983 \cite{UA1:1983crd,UA2:1983tsx,UA1:1983mne,UA2:1983mlz}. In 1999, 't Hooft and Veltman were awarded the Nobel prize for showing that the electroweak SM is a renormalizable theory.

\section{The Mass of the Higgs Boson} 

\begin{figure}
\centering
\vspace*{-1mm}
\includegraphics[scale=0.4]{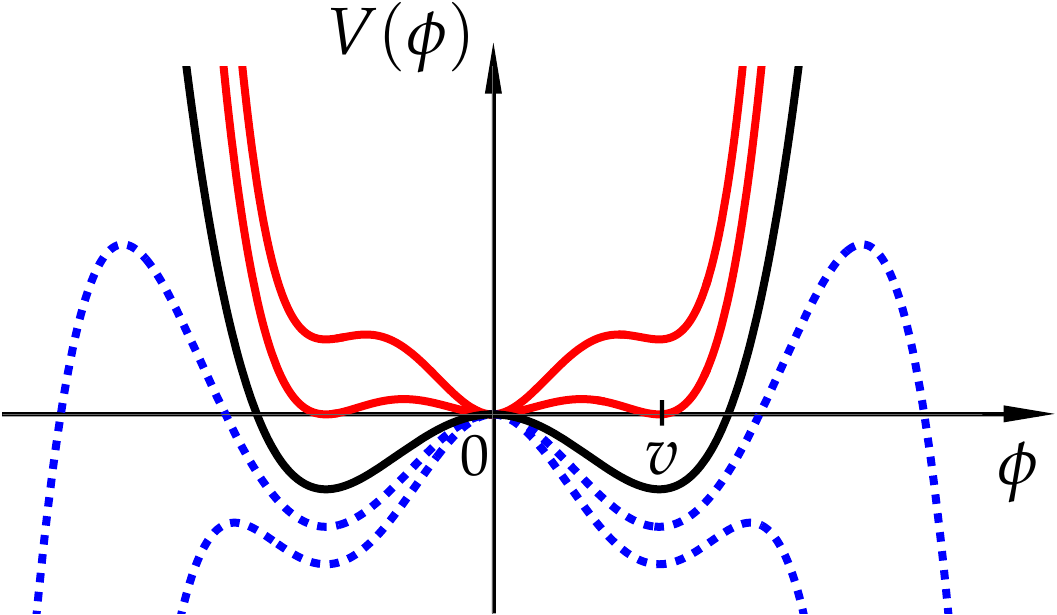}
\vspace*{-1mm}
 \caption{The possible destabilization of the electroweak vacuum. If the potential is such that $V(v)\ge V(0)$ SSB cannot occur (red solid lines); if the quartic coupling turns negative at large $\phi$, the vacuum is instable (dashed blue lines).} 
	\label{fig:unstable}
\vspace*{-1mm}
\end{figure}

As was discussed in the previous section, one of the pillars of the SM was the  spontaneous breakdown of the electroweak symmetry by the scalar field $\phi=v+H$ of eq.~(\ref{eq:Phiphi}) which bears the scalar potential given in eq.~(\ref{eq:Vscal}) and, for $\mu^2\!<\!0$, acquires a vev \mbox{$v = (\sqrt{2} G_F)^{-1/2} \approx\ 246\,$GeV}. However, radiative corrections must be included in the scalar potential, i.e. the loop contributions of gauge bosons, heavy fermions and the Higgs boson itself. The fundamental observation of Weinberg \cite{Weinberg:1976pe} and independently Linde \cite{Linde:1975sw} was that for field values close to zero, these contributions might make that the scalar potential is destabilized and its minimum is not at the value $V(v)$  but at $V(0)$ instead; see Fig.~\ref{fig:unstable}. Requiring the electroweak scale $v$ to be an absolute minimum, and not only a local one, would set a constraint on the Higgs self-coupling $\lambda$ and, hence, on the Higgs boson mass since we have $\lambda= M_H^2/2v^2$. This was a crucial observation since, at that time, no lower bound existed on $M_H$ and it was even possible that the particle could be massless and, hence,  extremely hard to observe experimentally. Let us briefly summarize how Weinberg presented the argument. 

The only corrections that are relevant in this context are those due to the massive gauge bosons whose couplings to the Higgs filed are given in terms of the \SU2 and \U1 coupling constants $g$ and $g'$. Indeed, at the time of the paper (1975),  there were only two generations of fermions  whose masses were too small for their contributions to matter.  Also, because they were looking for a lower bound on $M_H$, which was expected to be much smaller than $M_{W,Z}$, one can ignore the Higgs contributions as $\lambda$ is small. In this case, the scalar potential, when including the one-loop corrections from the $W^\pm/Z^0$ bosons, becomes
\begin{align}
\hspace*{-5mm}
V( \phi) \!=\! \frac{1}{2}\mu^2 \phi^2 \!+\! \lambda \frac{1}{4} \phi^4 \!+\! \frac{3}{64\pi^2} [2g^4\!+\!(g^2\!+\!g'^2)^2] \phi^4 \log \frac{ \phi^2}{\mu_R^2}\, , \label{eq:RGE}
\end{align}
where $\mu_R$ is a renormalization scale. For spontaneous symmetry breaking to occur, $V(\phi)$ should have a non-trivial local minimum at $\phi=\phi_0$ such that the following conditions are met 
\begin{equation}
\phi_0 \neq 0\, , \ \   \delta V/\delta \phi|_{ \phi=\phi_0}=0 \, , \  \ \delta^2V/\delta \phi^2 |_{ \phi=\phi_0}= M_H^2 \geq 0 \, . 
\end{equation}
The requirement that $V(\phi_0=v) <V (0)$, necessary to allow for a proper SSB, would then imply a bound on the Higgs mass,
\begin{eqnarray}
M_H^2 \geq \frac{3 \sqrt 2 G_F }{16 \pi^2} (2M_W^4+M_Z^4)\, ,   
\end{eqnarray}
where the couplings $g$ and $g'$ in eq.~(\ref{eq:RGE}) are expressed in terms of the $W^\pm,Z^0$ masses. This gives the value $M_H \gsim 7$ GeV.\footnote{In 1975, the value of $M_{W/Z}$ (or $\sin\theta_W$) were not known and, assuming $M_Z \gsim M_W$, the bound which was obtained by Weinberg is $M_H \gsim 3.72$ GeV.}

The contemporary way to look at this bound on $M_H$ from the stability of the electroweak vacuum is not to consider the infrared regime $\phi \to 0$ but, instead, the ultraviolet regime \mbox{$\phi\to M$} where $M$ is a high mass scale, e.g., the Planck scale $M_P$. 

Let us see how it is presented in modern language. The dominant effect of the  inclusion of the radiative corrections to the scalar potential, when summing the leading logarithms to all orders, would be to replace the two terms $\mu^2$ and $\lambda$ by running parameters evaluated at the relevant energy scale $Q$ of the field $\phi$. The evolution of the self-coupling $\lambda$ with $Q$, when including all relevant contributions, i.e. those of the $W^\pm/Z^0$ bosons, the fermions and the Higgs boson itself,  is approximately described by the renormalization group equation (RGE) 
\begin{align}
 \frac{\dd \lambda}{\dd \log Q^2}  \simeq & \frac{1}{16\pi^2} \left[
12 \lambda^2 + 
\frac{3}{16} \left(2 g^4+ (g^2+g'^{2})^2\right) -3\lambda_t^4 \right]\, ,   \label{eq:RGE2}
\end{align}
where we have taken into account only the contribution of the top quark with Yukawa coupling $\lambda_t= \sqrt 2 m_t/v$; the minus sign in front of it comes from Fermi statistics and plays a crucial role. Indeed, because the top quark is so heavy  one can ignore, this time, the contribution of the Higgs boson, assuming $\lambda\ll g,g',\lambda_t$. The solution to the previous RGE, taking again the weak scale as the reference point,  would be in this case, 
\begin{equation}
\lambda(Q^2)\!=\!\lambda(v^2)\!+\! \frac{1}{16 \pi^2} \left[ \frac{3}{16} \left(2 g^4\!+\! (g^2\!+\!g'^{2})^2 \right) \!-\! 3 \lambda_{t}^4 \right] \log \frac{Q^2}{v^2} \, . 
\end{equation}
If the coupling $\lambda$ (or $M_H$) is too small, the top quark contribution can be dominant and could  drive it to a negative value $\lambda(Q^2) <0$, leading to a scalar potential $V(Q)\! <\! V(v)$ \cite{Cabibbo:1979ay,Politzer:1978ic,Hung:1979dn,Sher:1988mj}. The vacuum is not stable and in fact not bounded from below; see dashed lines of Fig.~\ref{fig:unstable}. To have a well behaved vacuum, $M_H$ needs to satisfy the requirement 
\begin{eqnarray}
M_H^2 > \frac{3 \sqrt 2 G_F }{16 \pi^2}  \left[ 4 m_{t}^4 -(2M_W^4+M_Z^4) \right]\log \frac{Q^2}{v^2} \, .  
\end{eqnarray}
The constraint or lower bound on $M_H$ depends on the value of the cutoff scale $Q=\Lambda$ up to which the theory is expected to be valid. For $\Lambda \approx 1\,{\rm TeV}$ one has, in this first approximation (refinements are discussed below), $M_H \gsim 50\, {\rm GeV}$ while for $\Lambda \approx M_P \approx 2\cdot 10^{18}\,{\rm GeV}$ one would have $M_H \gsim 250~{\rm GeV}$.

\begin{figure}[!ht]
	\centering
\vspace*{-.1mm}
	\includegraphics[width=0.35\textwidth]{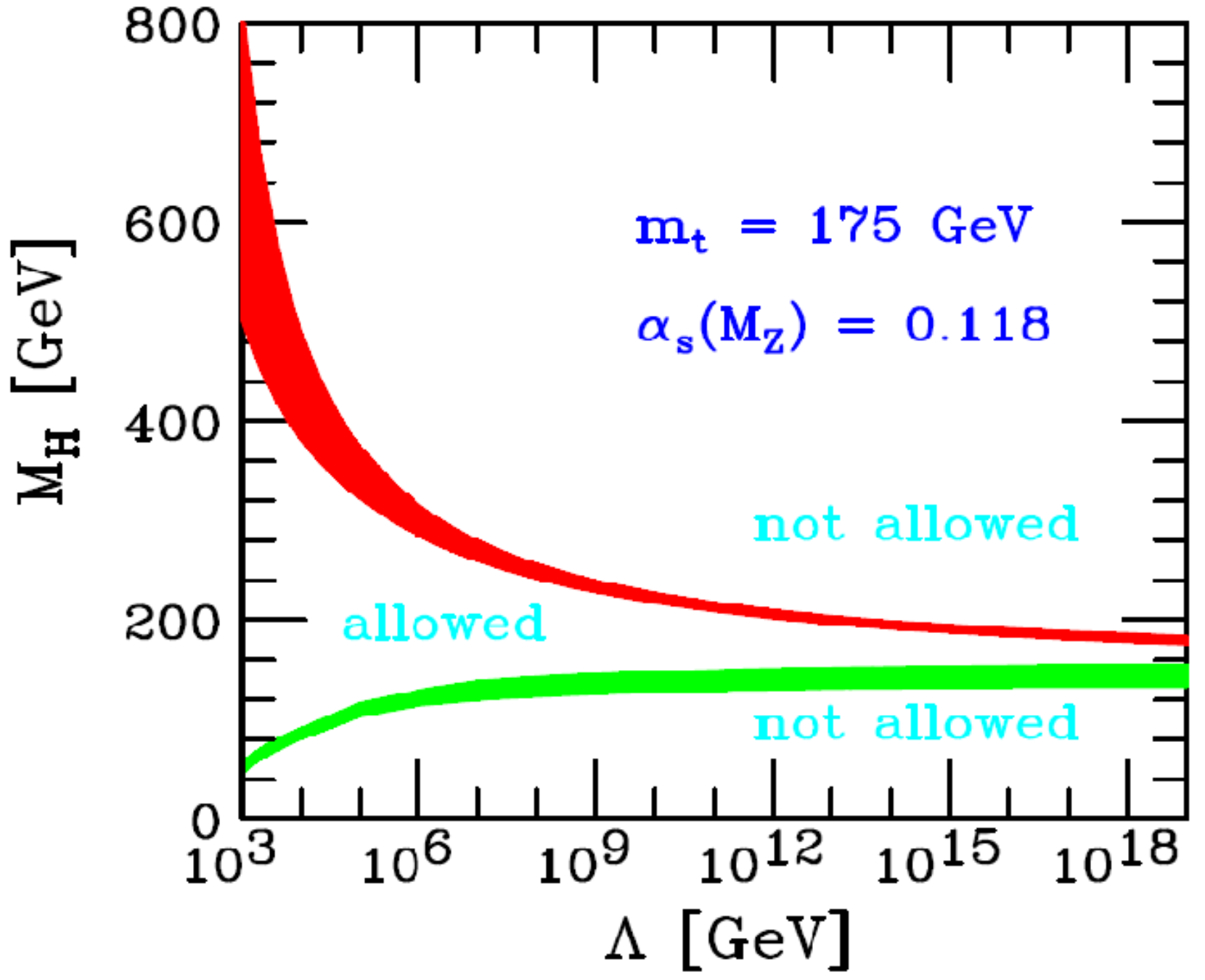}	
\vspace*{-1mm}
	\caption{Triviality (upper) bound and vacuum stability (lower) bounds on the Higgs mass as functions of the scale $\Lambda$ including experimental and theoretical uncertainties  (thickness of the bands) with the knowledge of 1995~\cite{Hambye:1996wb}.} 
	\label{fig:Vela-HR}
\vspace*{-2mm}
\end{figure}

A few remarks are interesting and enlightening at this stage. A first one is that, in parallel to the Linde-Weinberg lower bound, an upper bound of $M_H \lsim 1$ TeV was obtained from the requirement of perturbative unitarity in the scattering of the $W^\pm/Z^0$ bosons at high energy \cite{Dicus:1973gbw,Cornwall:1974km,Veltman:1976rt,Lee:1977eg}. At the same time, the coupling $\lambda$ becomes too large and perturbation theory unreliable for $M_H = {\cal O}(1\, {\rm TeV})$. In fact, $\lambda$ would then dominate the RGE of eq.~(\ref{eq:RGE2}) which would then have a simple solution 
\begin{equation}
\lambda (Q^2) = \lambda (v^2) \bigg[ 1 - \frac{3}{4\pi^2} \, \lambda (v^2) \,  \log \frac{Q^2}{v^2} \bigg]^{-1}  \, , 
\end{equation}
leading to the so-called triviality bound: for $Q^2\! \ll\! v^2$, $\lambda$ becomes extremely small and eventually vanishes, making the theory non-interacting and hence trivial \cite{Wilson:1973jj}. In the opposite high-energy limit, $Q^2 \! \gg v^2$, $\lambda$ grows and eventually becomes infinite at a point called the Landau pole  $\Lambda = v \, \exp \left ( {2 \pi^2}/{3\lambda} \right)$. To avoid this pole, one needs (after some refinements made e.g. in Ref.~\cite{Hambye:1996wb}) $M_H \lsim 200$ GeV for $\Lambda \approx M_P$, while for the lowest  cutoff $\Lambda = {\cal O}(M_H)$ one gets $M_H \lsim 1\,{\rm TeV}$. This upper bound was combined with the improved lower one from vacuum stability to obtain the famous Roman plot \cite{Cabibbo:1979ay} which, for three decades, cornered the Higgs mass; see a version from 1995 in Fig.~\ref{fig:Vela-HR}. 

A second remark is that the lower bound on $M_H$ can be relaxed if the vacuum is metastable~\cite{Degrassi:2012ry}.  Indeed, the effective potential can have a minimum which is deeper than the standard electroweak minimum if the decay of the latter into the former, e.g. via thermal fluctuations in the hot universe (which are subject to some cosmological assumptions) or quantum fluctuations at zero temperature (a tunneling which depends on  the lifetime of the instability of the vacuum compared to the age of the universe) is suppressed. In this case, a lower bound on $M_H$ follows from the requirement that no transition between the two vacua occurs and we always remain in the electroweak minimum. The obtained  lower bound is in general much weaker than in the case of absolute stability of the vacuum and even disappears if the cutoff of the theory is at the TeV scale.

Finally, to obtain a very precise lower bound on $M_H$, the condition of absolute stability of the electroweak vacuum should be derived including the state-of-the-art quantum corrections at next-to-next-to-leading order in perturbation theory. Presently, the full calculation is based on three main ingredients that have been obtained only a decade ago \cite{Degrassi:2012ry,Bezrukov:2012sa}:  the two-loop threshold corrections to the quartic coupling at the weak scale which involve the QCD and the Yukawa interactions, the three-loop leading contributions to the Higgs mass anomalous dimension and to the RGEs of Higgs self-coupling and the top quark Yukawa coupling, and the three-loop corrections to the beta functions of the three SM gauge couplings taking into account the two couplings above. Lastly, the most precise values of the Higgs mass, the top  mass and all the couplings including the strong one $\alpha_s$ should be used as inputs. 

\begin{figure}[!ht] 
	\centering
\vspace*{-2mm}
                \includegraphics[width=0.3\textwidth]{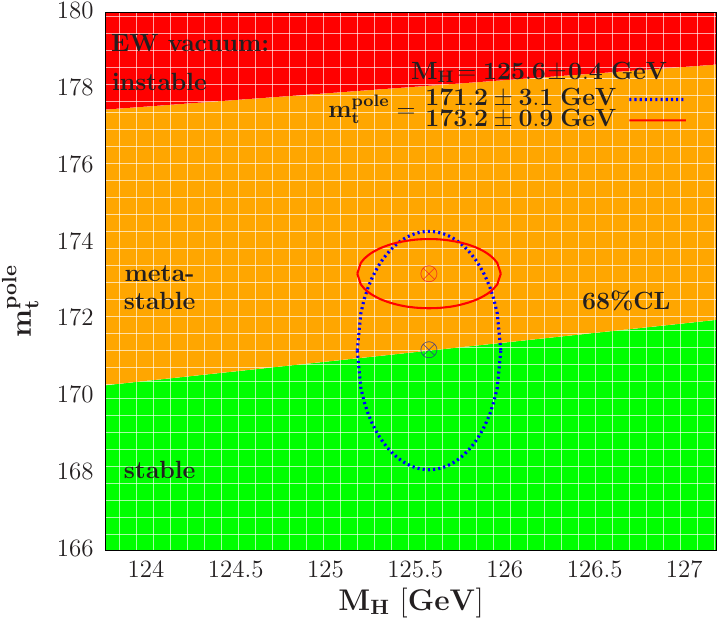}	  
\vspace*{-1mm}    
	\caption{The vacuum stability bound on the Higgs mass in the plane $[M_H, m_{t}]$. From Ref.~\cite{Alekhin:2012py} which appeared the next day of Higgs discovery. } 
	\label{fig:stability}
\vspace*{-1mm} 
\end{figure}

The outcome of the calculation is exemplified in the  plane $[M_H, m_{t}]$ in Fig.~\ref{fig:stability} taken from Ref.~\cite{Alekhin:2012py}, when the SM is extrapolated up to the Planck scale. As can be seen, for the measured Higgs and top quark masses, one is close to the critical boundary for vacuum stability. This has opened very interesting possibilities for new physics model building \cite{Degrassi:2012ry} and requires a more precise measurement of the top quark mass \cite{Alekhin:2012py}.  

\section{Flavor and CP-violation in extended Higgs models}

It was noticed rather early after Weinberg's historical paper that there is no fundamental reason, except for minimality, for the Higgs sector of the SM to consist of only one Higgs doublet.  It can be extended to contain any multiplet provided that the scalar potential is invariant under the ${\SU2_{L} \times \U1_{Y}}$ gauge symmetry. There were, however, two important phenomenological constraints that any extended Higgs sector must satisfy. 

A first one comes from the $\rho$ parameter which,  historically was used to measure the relative strength of the neutral to the charged weak currents at zero-momentum transfer in deep-inelastic neutrino-nucleon scattering \cite{Veltman:1977kh}. In the SM, it is given in terms of the $W^\pm/Z^0$ masses and the cosine of the Weinberg angle by $\rho=  M_W^2/(\cos^2 \theta_W M_Z^2)$ and, up to small radiative corrections,  $\rho \simeq 1$, as experimentally observed. This is a direct consequence of the choice for the representation of the Higgs field. Indeed, in a model which makes use of an arbitrary number of Higgs multiplets $\Phi_i$ with isospin $I^i$ and third component $I_3^i$, whose neutral states acquire vevs $v_i$, one obtains
\begin{eqnarray}
\rho = \frac{\sum_i \big[ I^i (I^i+1) -(I_3^i)^2\big] v_i^2}{2\, \sum_i (I_3^i)^2 v_i^2}\, . 
\end{eqnarray} 
It is equal to unity only for doublet and scalar singlet  fields. This is a consequence of the model having a custodial ${\SU2_{R}}$ global symmetry that is broken at the loop level when fermions of the same doublets are non-degenerate in mass and by the hypercharge group.\footnote{The main contribution to $\rho-1$ is due to the top-bottom isodoublet as the large mass splitting between the two quarks breaks the custodial symmetry and generates a contribution  which grows as the top mass squared. Another contribution, but logarithmic this time, is due to the Higgs boson, $\propto \frac{\alpha}{\pi} \log (M_H^2/M_W^2)$. It is the deviation $\rho-1$ that allowed to constrain the masses of the top quark and the Higgs boson before they were directly observed and with possible values close to those that have been measured. This made the triumph of the SM.} Any other Higgs representation, e.g. scalar triplets, would require fine-tuning of the Higgs vevs to satisfy this constraint. Hence, most extensions of the Higgs sector only involved additional scalar fields in doublets or singlets. 

A second extremely important constraint on extended Higgs sectors  stems from the absence of flavor changing neutral currents (FCNC) at tree-level. In the SM, FCNC are suppressed because of the exact electromagnetic symmetry and of the GIM mechanism \cite{Glashow:1970gm}, i.e. the fact that the $3\times3$ CKM matrix that turns the down-type current-eigenstate quarks into mass eigenstates is unitary, insuring that neutral currents are diagonal in both bases and satisfy the severe  experimental bounds. 

In extensions of the Higgs sector and, in particular, in models with additional Higgs doublets, this is in general not true as two or more Yukawa matrices for each fermion cannot be simultaneously diagonalized and FCNC associated with Higgs boson exchange naturally occur. In the seminal 1976 paper ``Natural conservation laws for neutral currents" \cite{Glashow:1976nt}, Glashow and Weinberg (and independently Paschos \cite{Paschos:1976ay}) proposed a theorem that solves the problem in a rather elegant way: tree-level Higgs mediated FCNCs are absent if all fermions of a given electric charge couple to not more than one Higgs doublet. 

In practice, this can be achieved by imposing additional discrete $\mathbb{Z}_2$ symmetries to the Higgs fields. The Yukawa interaction of the Higgs bosons to fermions are then severely constrained but they are not unique.  Let us briefly summarize the implications of this theorem in the important and widely studied case of a two-Higgs doublet model (2HDM) with fields $\Phi_1$ and $\Phi_2$ \cite{Aoki:2009ha,Branco:2011iw}. A  $\mathbb{Z}_2$ symmetry is introduced under which they are even or odd: $\Phi_1 \mapsto \Phi_1$  and $\Phi_2 \mapsto -\Phi_2$. The most general Yukawa Lagrangian involving it can be then written as
\begin{align}
{\mathcal L}_\text{\rm Yuk} ^\text{2HDM} =
&\!-\!{\overline Q}_L \lambda_u \widetilde{\Phi}_u u_R^{}
\!-\!{\overline Q}_L \lambda_d \Phi_d d_R^{}
\!-\!{\overline L}_L \lambda_\ell \Phi_\ell \ell_R^{}+ \text{h.c.}\, ,
\end{align}
where, using the notation of the first generation, $\Phi_f$ with $f\!=\!u,d,\ell$ is either $\Phi_1$ or $\Phi_2$. The four independent $\mathbb{Z}_2$ charge assignments on quarks and charged leptons, leading to four model types, are summarized in Table  \ref{tab:Z2}. In the Type-I model,  all charged fermions obtain their masses from the $\Phi_2$ state, while in Type-II, the masses of up-type quarks come from $\Phi_2$ and those of down-type quarks and leptons come from the $\Phi_1$ field. Type-X and Y are when the lepton $\ell$ couples differently from the quark $d$. The celebrated minimal supersymmetric extension of the SM or MSSM, is a 2HDM of Type-II \cite{Gunion:1989we,Djouadi:2005gj}. 

\begin{table}[!ht]
\begin{center}
\vspace*{-2mm}
\renewcommand{\arraystretch}{1.15}
\begin{tabular}{|c||c|c|c|c|c|c|}
\hline & $\Phi_1$ & $\Phi_2$ & $u_R^{}$ & $d_R^{}$ & $\ell_R^{}$ & $Q_L$/$L_L$ \\  \hline
{\small Type-I}  & $+$ & $-$ & $-$ & $-$ & $-$ & $+$ \\
{\small Type-II} & $+$ & $-$ & $-$ & $+$ & $+$ & $+$ \\
{\small Type-X}  & $+$ & $-$ & $-$ & $-$ & $+$ & $+$ \\
{\small Type-Y}  & $+$ & $-$ & $-$ & $+$ & $-$ & $+$ \\
\hline
\end{tabular}
\end{center}
\vspace*{-3mm}
\caption{Charge assignments of the discrete $\mathbb{Z}_2$ symmetry in the 2HDM.} \label{tab:Z2}
\vspace*{-2mm}
\end{table}

The couplings of the fermions to the five 2HDM scalar states, two CP-even $h$ and $H$, a CP-odd $A$ and two charged $H^\pm$ bosons are given by the Lagrangian (we exhibit only those in the neutral case, the couplings in the charged case are similar)
\begin{equation}
\mathcal{L}_{\rm Yuk} =\Sigma_f (m_f/v)\! \times \!  [g_{hff} h \bar f f+g_{Hff} H\bar f f- i g_{Aff} A  \bar f \gamma_5 f] \, . 
\end{equation}
The couplings $g_{\phi ff}$ normalized to the SM-Higgs couplings are given in terms of two angles: $\beta$ from the ratio of the two Higgs vevs  $\tan\beta=v_2/v_1$ and the mixing angle $\alpha$ that diagonalizes the CP-even Higgs mass matrix. Their values in the so-called alignment limit when $h$ is SM-like, $g_{hff}\!=\!1$, depend only on $ \tan\beta$ as $\alpha\! \to\! \beta\!-\!\frac{\pi}{2}$ and are given by $g_{Hff}= g_{Aff} = \xi_f$ with values in the four types of 2HDMs are summarized in Table \ref{2hdm_type}.

\begin{table}[h!]
\renewcommand{\arraystretch}{1.2}
\vspace*{-1mm}
\begin{center}
\begin{tabular}{|c|c|c|c|c|}
\hline
~~~~~~ &  ~~{\small Type-I}~~ & ~~{\small Type-II}~ & {\small Type-X} & {\small Type-Y} \\ \hline\hline 
$\xi_{u}$ & $\cot\beta$ & $\cot\beta$ &$\cot\beta$ & $\cot\beta$\\ \hline
$\xi_{d}$ & $-\cot\beta$ & ${\tan\beta}$ & $-\cot\beta$ & ${\tan\beta}$ \\ \hline
$\xi_{\ell}$ & $-\cot\beta$ & ${\tan\beta}$ & ${\tan\beta}$ & $-\cot\beta$ \\ \hline
\end{tabular}
\vspace*{-1mm}
\caption{Possible values, in the alignment limit $\beta \!-\! \alpha \rightarrow \frac{\pi}{2}$, of the $\xi_f$ parameters describing the couplings of the extra Higgs bosons to the SM fermions.}
\label{2hdm_type}
\end{center}
\vspace*{-3mm}
\end{table}

The Type-II and X scenarios feature enhanced $H/A/H^\pm$ couplings to the isospin down-type  fermions for large values of $\tan\beta$. This has important consequences in many cases. 

We should further note that the  $\mathbb{Z}_2$ symmetry also constrains the parameters, i.e. the mass terms and the quartic couplings, of the scalar potential of the model. For instance, using the reflection symmetry and simultaneously requiring CP-invariance, the 2HDM scalar potential would read:
\begin{align}
 V_{2HDM} =& m_{11}^2 \Phi_1^\dagger \Phi_1+ m_{22}^2 \Phi_2^\dagger \Phi_2 - m_{12}^2  (\Phi_1^\dagger \Phi_2 + {\rm h.c.} )  \nonumber \\ & +\tfrac12{\lambda_1} ( \Phi_1^\dagger \Phi_1 )^2 +\tfrac12{\lambda_2} ( \Phi_2^\dagger \Phi_2 )^2 
+\lambda_3 (\Phi_1^\dagger \Phi_1) (\Phi_2^\dagger \Phi_2)  \nonumber \\ & +\lambda_4 (\Phi_1^\dagger \Phi_2 ) (\Phi_2^\dagger \Phi_1) +\tfrac12 {\lambda_5} \big[\,   (\Phi_1^\dagger \Phi_2 )^2 + {\rm h.c.} \, \big] \, .
\label{eq:V2HDM}
\end{align}
As can be seen, it has only 8 free parameters as two couplings; two have been removed and we have allowed for the operator $\Phi_1^\dagger \Phi_2+\mbox{h.c.}$ which softly breaks the $\mathbb{Z}_2$ symmetry. 

Besides this work, which played a crucial role in selecting the phenomenologically viable scalar sectors in new physics model-building, Weinberg made other important contributions related to the impact of the Higgs bosons in flavor physics. For instance, in an other very important paper \cite{Weinberg:1976hu}, he noted that in extensions with three or more Higgs doublets, the scalar potential which is invariant under the discrete symmetries that forbid FCNC could lead to the violation of the CP symmetry.

In 1977, he suggested a three Higgs-doublet model (3HDM) \cite{Weinberg:1976hu}, a $\mathbb{Z}_2 \times  \mathbb{Z}_2 \times  \mathbb{Z}_2$ symmetry under the separate reflections of the Higgs doublets $\Phi_i \mapsto - \Phi_i$, as well as appropriate transformations for the quark fields, that strongly reduce the number of free parameters of the theory and ensure the absence of FCNC.  In the scalar potential of the model, a generalization of that of the 2HDM in eq.~(\ref{eq:V2HDM}),  one could include three additional terms
\eq{
V_{\rm 3HDM} \supset
\sum_{i<j=1}^3 \ d_{ij}\, \e^{\ii\theta_{ij}} (\Phi_i^\dagger\Phi_j) +{\rm h.c.}\, , 
}
which can be complex. Weinberg remarked that, in general, one cannot rotate away simultaneously the three phases $\theta_{ij}$ and, hence, such a theory can explain in a natural way the violation of the CP symmetry. When the paper was written, only the first two generations of fermions were present. These had very small Yukawa couplings and, thus, allowed  a CP-violation with the correct strength.\footnote{These 3HDMs are now being discussed e.g. in the context of Dark Matter particles whose interactions with the SM are mediated by the Higgs-portal \cite{Arcadi:2019lka}.}  

More than a decade later, Weinberg returned to the subject of Higgs sector effects on CP-violating observables and made two important contributions \cite{Weinberg:1989dx,Weinberg:1990me}. In Ref.~\cite{Weinberg:1989dx}, he introduced a dimension-6 CP-violating chromo-electric dipole moment mediated by the QCD vertex involving three gluons,  
\begin{equation} 
{\cal O}_W= f^{abc} \epsilon^{\alpha\beta\gamma\delta}  G_{\mu \alpha}^a G_{\beta \gamma}^b G_\delta^{c\mu} / 3! \, , 
\end{equation}
where $f^{abc}$ is the \SU3 structure constant with color indices $a,b,c$ and $\epsilon$ the anti-symmetric tensor; $G_{\mu \nu}$ is the gluon field strength. This CP-odd  interaction is generated by a heavy quark loop in which Higgs bosons are exchanged and gives rise to a potentially large CP-violating contribution to the neutron electric dipole moment from the Higgs sector. In the SM, the contribution due to the CP-violating phase of the CKM matrix is negligible as it is suppressed by the GIM mechanism. However, the CP-violating contributions can be large in extended Higgs sectors as he himself has shown to be the case in 2HDMs (including those appearing in supersymmetric models) in  Ref.~\cite{Weinberg:1990me}. 

\section{Weinberg's Legacy}

Steven Weinberg is among the fathers of particle physics. In his historical ``model of leptons" of 1967, he initiated the modern way to view and to practice it by introducing its two cornerstones. The first one is the principle of unification using gauge theories which he successfully used to fuse the electromagnetic and weak interactions under the banner of the ${\SU2_{L} \times \U1_{Y}}$ symmetry group. This paved the way to the idea of unification of all  forces, a century after the fortunate merging of electricity and magnetism  into electromagnetism by Maxwell and half-a-century after the failed attempt to unify it with gravitation by Einstein and others.  The second cornerstone is the usage of the mechanism of spontaneous symmetry breaking, i.e. the fact that fields and interactions may obey these symmetries while the vacuum does not, to  generate particle masses. This mechanism is at the heart of electroweak unification but is employed in large variety of areas, including grand unified theories.  

The model was a curiosity when it was first put forward (even Weinberg was not really satisfied with it as it had a large number of arbitrary parameters) but it soon became clear that it gave the correct explanation for the weak interactions. This  occurred, in particular, after a few breakthroughs that came about in the early seventies. From the theoretical side, as discussed earlier, it got a decisive support when 't Hooft and Veltman proved it to be renormalizable. Shortly after, the neutral currents and, hence, indirectly the $Z^0$ boson that mediates them as predicted in the model, were discovered at CERN. In parallel, a revolution took  place in the description of the strong interactions. From the experimental side, quarks became a reality \cite{Bloom:1969kc,Breidenbach:1969kd} and the missing charm-quark was found \cite{E598:1974sol,SLAC-SP-017:1974ind}. From the theoretical side, this period witnessed the birth of modern QCD: the idea of asymptotic freedom \cite{Gross:1973id,Politzer:1973fx} allowed the utilization of perturbation theory for the strong interactions at high enough energy scales and the non-abelian ${\SU3_c}$ gauge theory description for it became absolutely transparent. 

\begin{figure}[!ht]
\vspace*{-1mm}
 \centering
\includegraphics[width=0.32\textwidth]{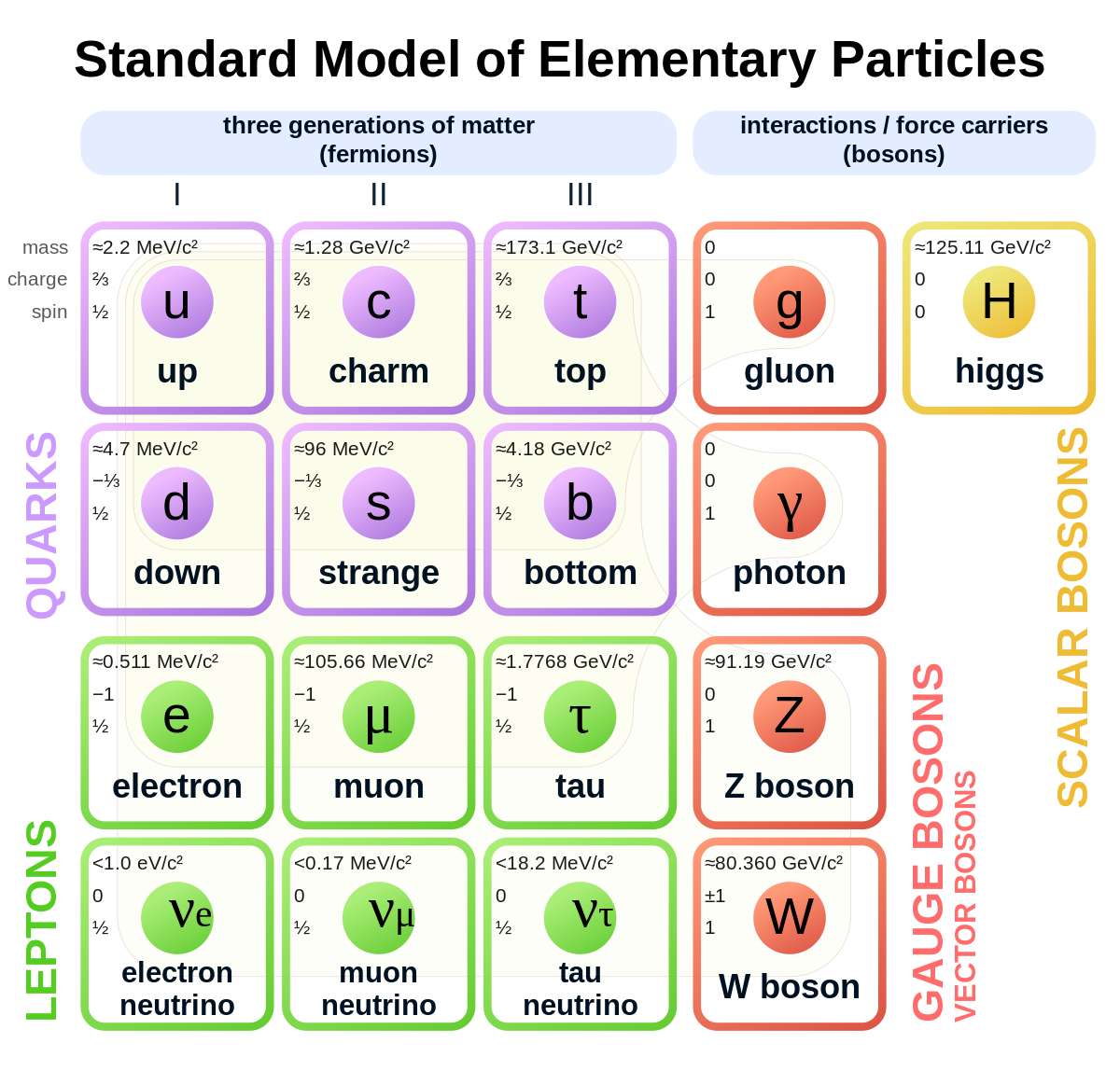}
\vspace*{-2mm}
\caption{The particle content of the Standard Model.} 
	\label{SM-content} 
\vspace*{-2mm}
\end{figure}

At the end of 1973, it was obvious that the three forces of Nature that are relevant in laboratory experiments can be described in a unified framework: a quantum field theory based on the gauge symmetry group ${\SU3_{c} \times \SU2_{L} \times \U1_{Y}}$ with the electroweak force spontaneously broken down to the electromagnetic ${\U1_{Q}}$ symmetry. Hence, all matter particles, the spin-$\frac12$ fermions, interact by exchanging vector particles, the spin-1 gauge bosons, and their masses are generated with their interaction with the spin-0 Higgs particle. This picture was strengthened by the discovery of the gluon  \cite{TASSO:1979zyf} in 1979 at DESY and the ${W^\pm}$ and $Z^0$ bosons \cite{UA1:1983crd,UA2:1983tsx,UA1:1983mne,UA2:1983mlz} at CERN in 1983. By the end of the 20th century, all particles of the SM were discovered, including a third generation of fermions, with the exception of the most singular one, the Higgs boson. The set of all particles present in the SM is shown in Fig.~\ref{SM-content}. 

\begin{figure}[!ht] 
\vspace*{-.4cm}
\begin{center} 
\begin{tabular}{ll}
\hspace*{-.2mm}
\begin{minipage}{12.cm}
\includegraphics[width=0.35\textwidth]{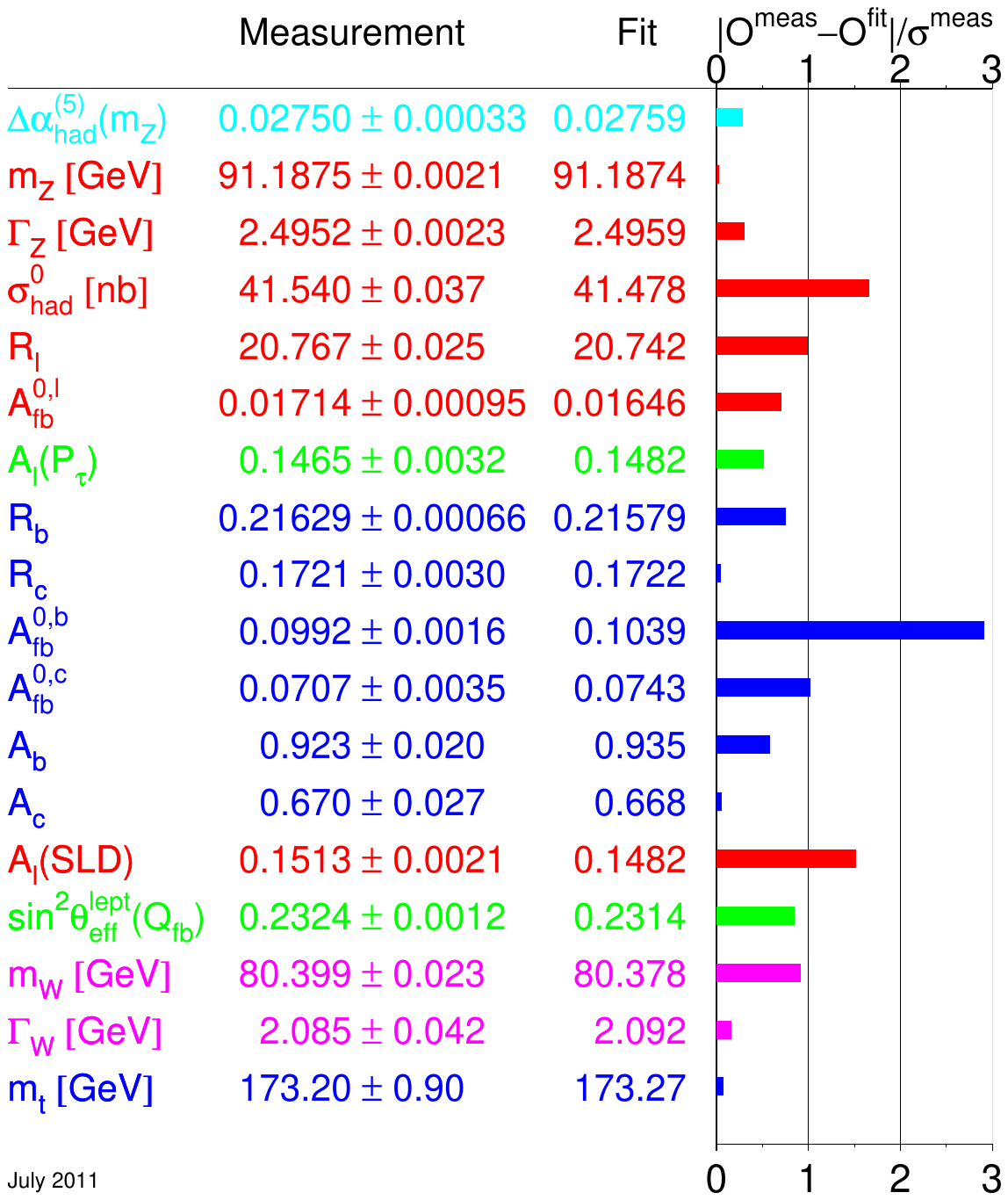}
\end{minipage} 
& \hspace*{-8.7cm}
\begin{minipage}{15cm}
\vspace*{-5mm}
\includegraphics[width=0.33\textwidth]{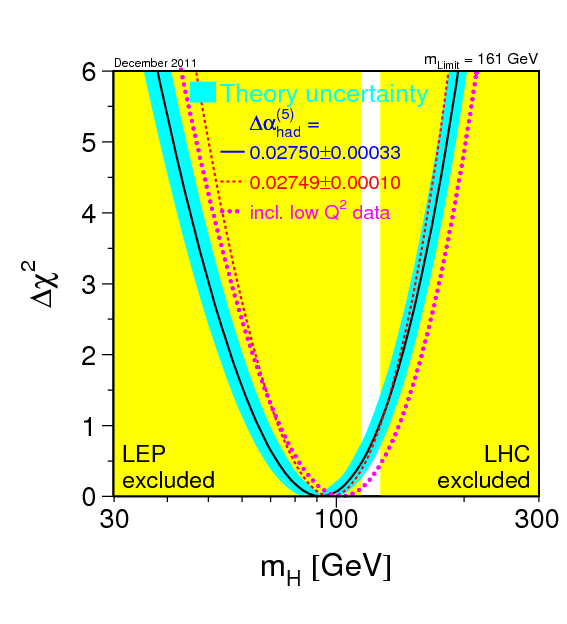} 
\end{minipage}
\end{tabular}
\end{center}
\vspace*{-1.3cm}
\caption[]{Left: precision measurements of the SM parameters at various experiments (mainly LEP) and comparison with the  theoretical  predictions. Right: global fit of the SM parameters and prediction on the mass of the Higgs boson in 2011, just before the discovery of the particle. From Ref.~\cite{ParticleDataGroup:2012pjm}.} 
\label{fig:blueband}
\vspace*{-.2cm}
\end{figure}

In the last decade of the twentieth century, high–precision measurements were carried out at LEP, SLC, the Tevatron and elsewhere. These tests, performed at the per mille level of accuracy or more, have probed the quantum corrections of the theory and the structure of the local gauge symmetry, providing a decisive confirmation that the SM is indeed the correct framework at energies close to the electroweak scale. This is shown in the left panel of Fig.~\ref{fig:blueband} where a large set of measurements and their impressive  accord with the SM are listed.  Alas, these experiments failed to observe directly the Higgs boson and could only set a lower bound of $114~{\rm GeV}$ on it mass, the only unknown parameter of the model. Nevertheless, by scrutinizing its quantum effects on the various precisely measured observables, an upper bound of $M_H \lsim 160$ GeV (125 GeV) was set at the 95\% (68\%) confidence level; see the right panel  of Fig.~\ref{fig:blueband}. 

\begin{figure}[!ht] 
\vspace*{-.2cm}
 \centerline{\hspace*{-2mm}
\includegraphics[width=0.28\textwidth]{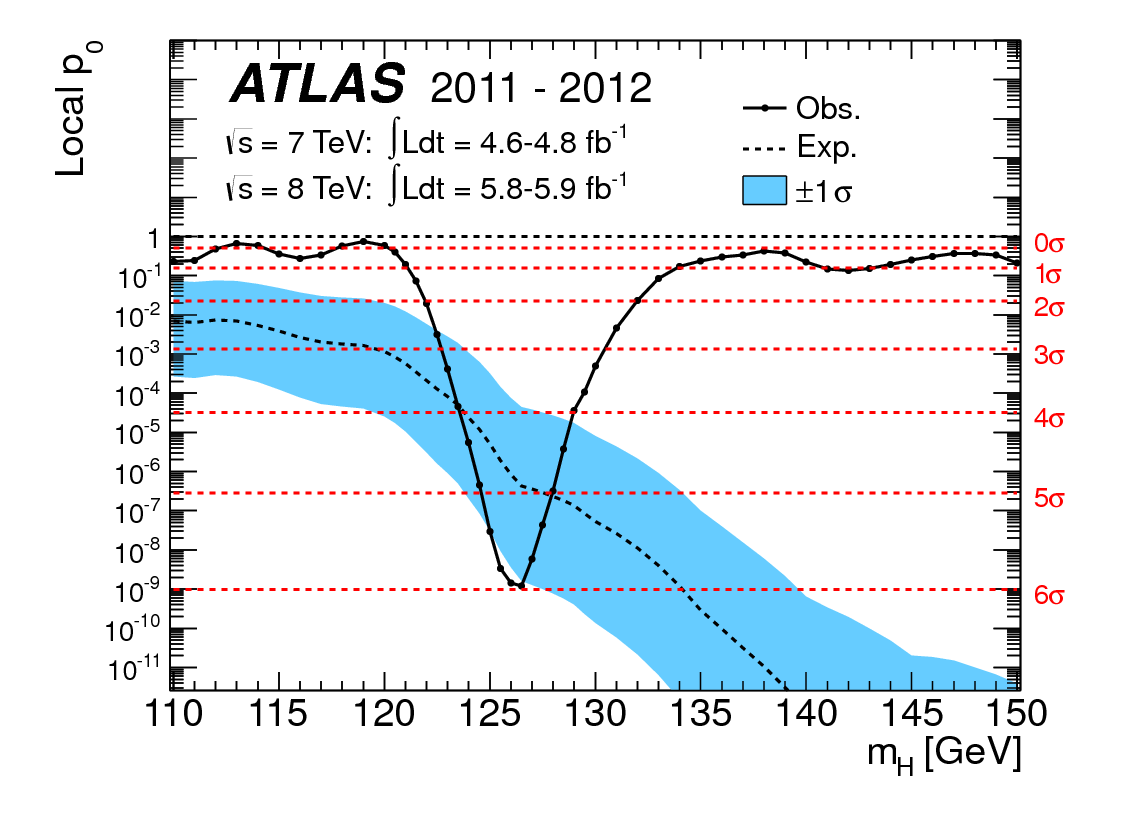}\hspace*{-2mm}
\includegraphics[width=0.222\textwidth]{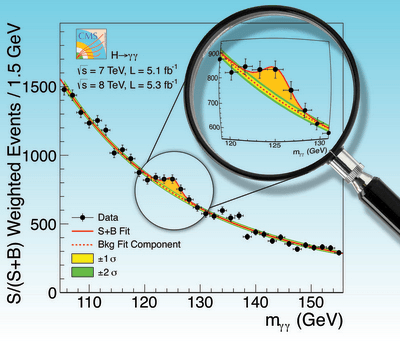} 
}
\vspace*{-.2cm}
\caption[]{The discovery plots of the SM Higgs boson in July 2012 at CERN when all data collected at two different c.m. energies were combined. Left: the likelihood of the discovery ($p$-values) as a function of $M_H$ in ATLAS \cite{Aad:2012tfa}. Right: the excess of the signal over the continuum background as a function of $M_H$ in the Higgs decay to two photons search channel in CMS \cite{Chatrchyan:2012xdj}.} 
\label{fig:discovery}
\vspace*{-.1cm}
\end{figure}

As it was discussed before, besides these experimental constraints, theoretical bounds on $M_H$ were also made available.  In addition to the lower bound from vacuum stability initiated by Weinberg and Linde and which excluded $M_H$ in the vicinity of 100 GeV,  an upper bound of ${\cal O}(1\, $TeV) was derived from the requirement of unitarity, triviality and perturbativity of the theory. This made it clear that the next generation of colliders after LEP, the LHC at CERN (as well as the late SSC in the US), were no-loose machines: either they would discover the Higgs boson or they should observe new phenomena connected with the new physics that plays its role. This conclusion was confirmed by the spectacular and historical discovery of the particle at CERN \cite{Aad:2012tfa,Chatrchyan:2012xdj} in July 2012 at precisely the mass,  $M_H\!=\!125$ GeV, predicted  by high-precision electroweak measurements (at 68\% confidence);  see Fig.~\ref{fig:discovery}. This completed the particle spectrum in the SM and marked another triumph.  

The discovery did not come as a surprise but was a coronation. As a matter of fact, soon after the discovery of the $W^\pm$ and $Z^0$ bosons in the early 1980's, probing the electroweak symmetry breaking mechanism became the dominant theme of elementary particle physics and the observation of its relic, the Higgs boson, the Holy Grail of high–energy colliders. 

Its detection required a gigantic effort from both experiment and theory. Indeed, besides the purely theoretical work of  constraining its mass, the complete ascertainment of the phenomenological profile  of the particle was of utmost importance for its unequivocal observation. The determination of its main production processes, its most important decay modes and the means and techniques by which it could be actually detected in experiment started in the mid-1970 with the work of e.g. Ref.~\cite{Ellis:1975ap}. It rapidly evolved and, by the end of the 1980's, a first complete guide for Higgs hunters \cite{Gunion:1989we} became available. It took two other decades to promote this knowledge to include the most important higher order effects and corrections and obtain a rather precise anatomy of the particle as summarized e.g. in Ref.~\cite{Djouadi:2005gi} and, just before its discovery, in Ref.~\cite{LHCHiggsCrossSectionWorkingGroup:2011wcg}. When the LHC started operating at around 2010, all the elements were assembled in order to leave no way of escape to the particle.  

\begin{figure}[!ht] 
 \centerline{
\includegraphics[width=0.31\textwidth]{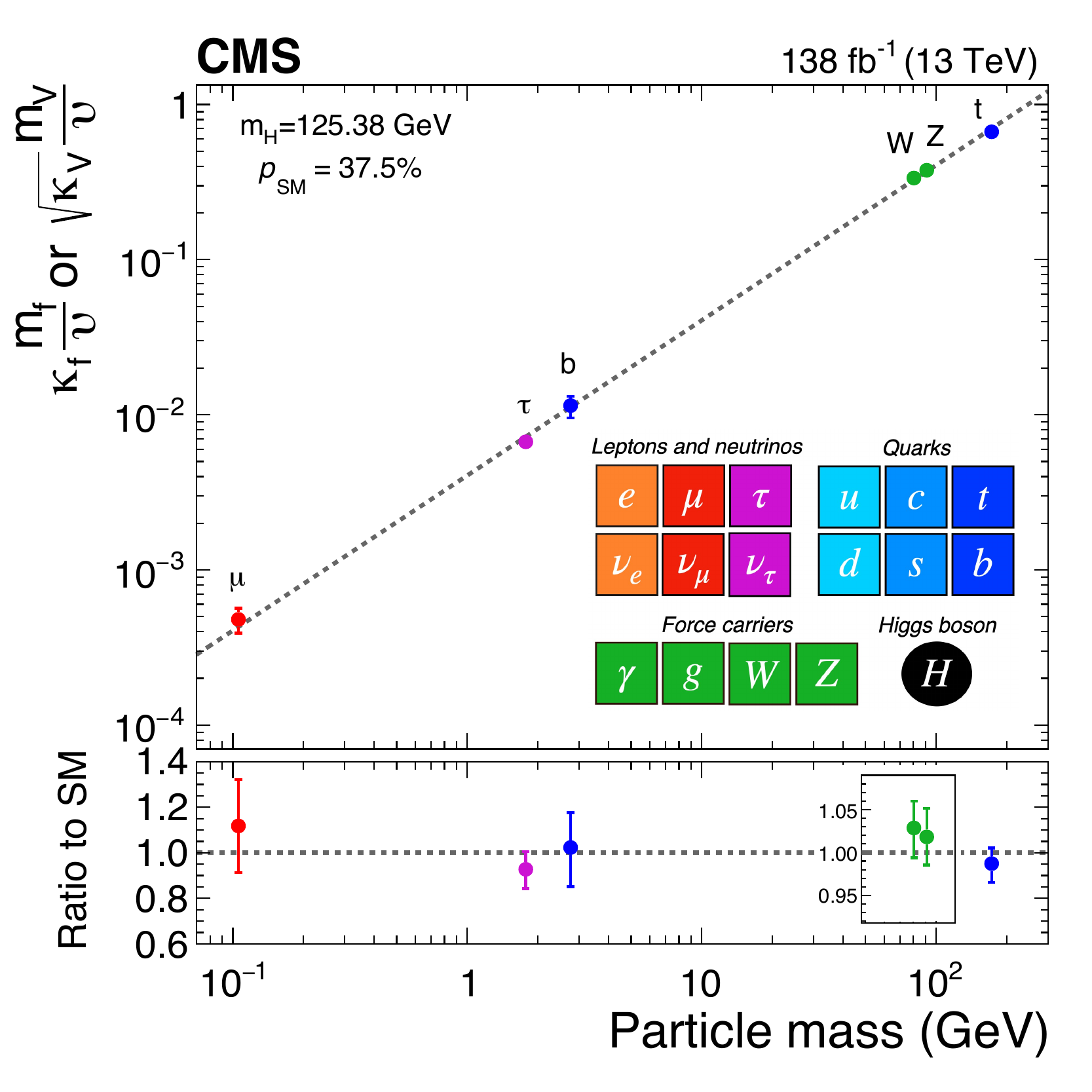} 
}
\vspace*{-.3cm}
\caption[]{The Higgs couplings to fermions and gauge bosons as functions of their masses relative to the values expected in the SM  as measured in 2022 at the LHC by the CMS collaboration \cite{CMS:2022dwd}. ATLAS obtains similar results \cite{ATLAS:2022vkf}.} 
\label{fig:Hcouplings}
\vspace*{-.3cm}
\end{figure}

Nevertheless, observing the Higgs boson was only a first part of the contract, as it is of equal importance to perform measurements of its basic properties in order to establish the exact nature of electroweak symmetry breaking and to achieve a more fundamental understanding of the phenomenon. In particular, besides verifying its spin and CP quantum numbers, it is of utmost importance to measure the couplings of the Higgs boson to the fermions and gauge bosons (as well as its self-coupling which, however, needs a wealth of more data than that available today) and check that they are indeed proportional to the particle masses, as initially suggested by Weinberg. A vast campaign has been launched at the LHC in order to precisely measure all the relevant Higgs decay and production rates and to extract  from them the Higgs couplings. The outcome of this effort is summarized in the spectacular plot shown in Figure \ref{fig:Hcouplings}, which demonstrates that these couplings are, at the 10\% level, indeed SM-like and are not altered by any other physics effect.  

This remarkable result is leaving the particle physics community with mixed feelings. On the one hand, one should be delighted to have a quantum field theory that explains most of the experimental data and the phenomena known to date. On the other hand, the theory is not satisfactory and is believed to be only effective and a low energy manifestation of a more fundamental one. Unfortunately, there is no sign of this new physics beyond the SM, neither indirectly through the measurements of the Higgs properties nor directly, by the direct production of new particles at the LHC. This is precisely the nightmare scenario that frightened people before the start of the LHC. 

As a matter of fact, and as Weinberg himself pointed to in his seminal 1967 paper, the SM is far from being perfect in many respects. It does not explain the proliferation of fermions and the large hierarchy in their mass spectra and does not say much about the small neutrino masses.\footnote{Weinberg was fascinated by the fermion mass problem and,  e.g.,  in one of his last papers \cite{Weinberg:2020zba} he contemplated models (that he himself considered to be still unrealistic) in which the masses of the third generation quarks and leptons arise at tree-level while those of the second and first generations are generated at the one- and two-loop levels respectively. For the neutrino masses, see below.} The SM does not unify in a satisfactory way the electromagnetic, weak and strong forces, as one has three different symmetry groups with three coupling constants which shortly fail to meet at a common value during their evolution with the energy scale (see below); it also ignores the fourth force, gravitation. Furthermore, it does not contain a particle that could account for the cosmological dark matter and fails to explain the baryon asymmetry in the Universe. 

In fact, Weinberg made major contributions in this context. In Ref.~\cite{Weinberg:1979sa},  he introduced  the only  (lepton-number violating) dimension-5 operator from SM fields, $\Lambda^{-1}\overline{L}\otimes\Phi\otimes\Phi\otimes L^c$, involving the Higgs and the lepton doublets, that upon SSB provides naturally  small masses to (Majorana) neutrinos for a large cutoff scale $\Lambda$. The three possible ways to complete the theory in the ultraviolet, i.e. the possible contractions of heavy fields with these ones, are a heavy neutrino singlet, a scalar triplet or a fermion triplet, corresponding to the so-called type I, II and III seesaw mechanisms, respectively; see e.g. Ref.~\cite{CentellesChulia:2018gwr}.

Another major work is about lepton and baryon numbers \cite{Weinberg:1979sa} that are good accidental global symmetries of the SM but are expected to be violated at high energies, a necessary ingredient to generate the observed baryon asymmetry of the Universe. Weinberg pioneered studies about this violation through higher-dimensional operators and as mentioned earlier, he e.g. introduced the lowest dimension-5 operator (called Weinberg operator) which violates lepton number by two units and gives rise to the small neutrino masses. Baryon number can be violated in one unit by four-fermion operators of dimension 6. These would lead to proton decay, the observation of which triggered a large and long lasting experimental programme. 

In fact baryon number violation and thus, proton decay, occurs naturally in Grand Unified Theories (GUT) since quarks and leptons belong to the same representation of the unifying group and there exist super-heavy gauge bosons (the leptoquarks) that will connect them. This was sought by Georgi, Quinn and Weinberg as early as 1974
\cite{Georgi:1974yf}. In this pioneering  paper, they considered the merging of the $\SU3_{c}$, $\SU2_{L}$, $\U1_{Y}$ groups that form the SM into a single GUT group like SU(5) \cite{Georgi:1974sy} with one coupling constant $g_{\rm GUT}$. By requiring the unification of the three gauge couplings at a high scale, estimated to be $M_{\rm GUT} = {\cal O}(10^{16})\,$GeV with the existing knowledge, they were able to predict a good value for the weak mixing angle, $\sin^2\theta_W \approx 0.2$, not too far from its true value (a ``tour de force"!). This program of   complete unification of the three forces became a very active field and we know now that, in the SM, the three gauge couplings fail to meet at a single point and the possible GUT scale is too low to prevent proton decay. New physics is thus required for a realistic unification.

Nonetheless, the main problem that called for new physics beyond the SM  was presumed to be related to the special status of the Higgs boson that has a mass which, contrary to fermion and gauge boson masses that are protected by chiral or gauge symmetries, cannot be preserved from quantum corrections that are quadratic \cite{Weisskopf:1939zz} in the new physics scale that serves as a cutoff  which, if too large, will drive $M_H$ to very large values, ultimately to the Planck scale instead of $M_H = {\cal O}(100\,{\rm GeV})$. Thus, the SM cannot be extrapolated to a scale beyond ${\cal O}(1\,{\rm TeV})$ where new phenomena should emerge. This is the reason why something new was expected  to manifest itself at the LHC.

There are three grand avenues for the many new physics scenarios that have been proposed to solve this SM problem of hierarchy of scales. First, there are theories with extra space-time dimensions that emerge at the TeV scale at which gravitational interactions come into play making that the cutoff of the theory is $\Lambda\!=\! {\cal O}({\rm few~TeV})$ and not $\approx\!  2\! \cdot\!  10^{18}$ GeV. A second avenue are composite models inspired from strong interactions, but occurring at the TeV and not at the 100 MeV scale, in which the Higgs boson is not a fundamental spin-zero particle but a bound state of some heavy fermions. Finally, the option that emerged in the most spectacular way is supersymmetry in which a new partner to each SM particle, but with a spin differing by $\frac12$, is postulated and exactly cancels the intolerable quadratic divergences in the Higgs mass; again, these new particles should not be much heavier than ${\cal O}(1{\rm TeV})$ not to spoil this miraculous compensation.  These extensions could also resolve other shortcomings of the SM such as providing the tiny missing contributions that would allow to unify the three SM couplings into the single coupling of a GUT, as seen earlier, and supplying a good candidate for the missing Dark Matter in the Universe. 

As already mentioned in the introduction, Weinberg made groundbreaking contributions in the foundation and development of these three new physics directions. For instance, he pioneered the possibility that the masses of the intermediate vector bosons arise from a dynamical symmetry breaking mechanism and posited the existence of extra-strong interactions, called Technicolor later, whose dynamics are like QCD but at a scale of 1 TeV \cite{Weinberg:1975gm}. He also made major contributions in establishing supersymmetry as the best beyond the SM candidate; he instigated phenomenological studies of its particle spectrum \cite{Weinberg:1981wj,Farrar:1982te} and the fascinating possibility that gravity could be the mediator of supersymmetry-breaking \cite{Hall:1983iz} in GUTs. He even  did some early work on extra dimensions; see e.g. Ref.~\cite{Weinberg:1983xy}. 

\begin{figure}[!ht] 
\vspace*{-.3cm}
\centerline{ \includegraphics[width=0.47\textwidth]{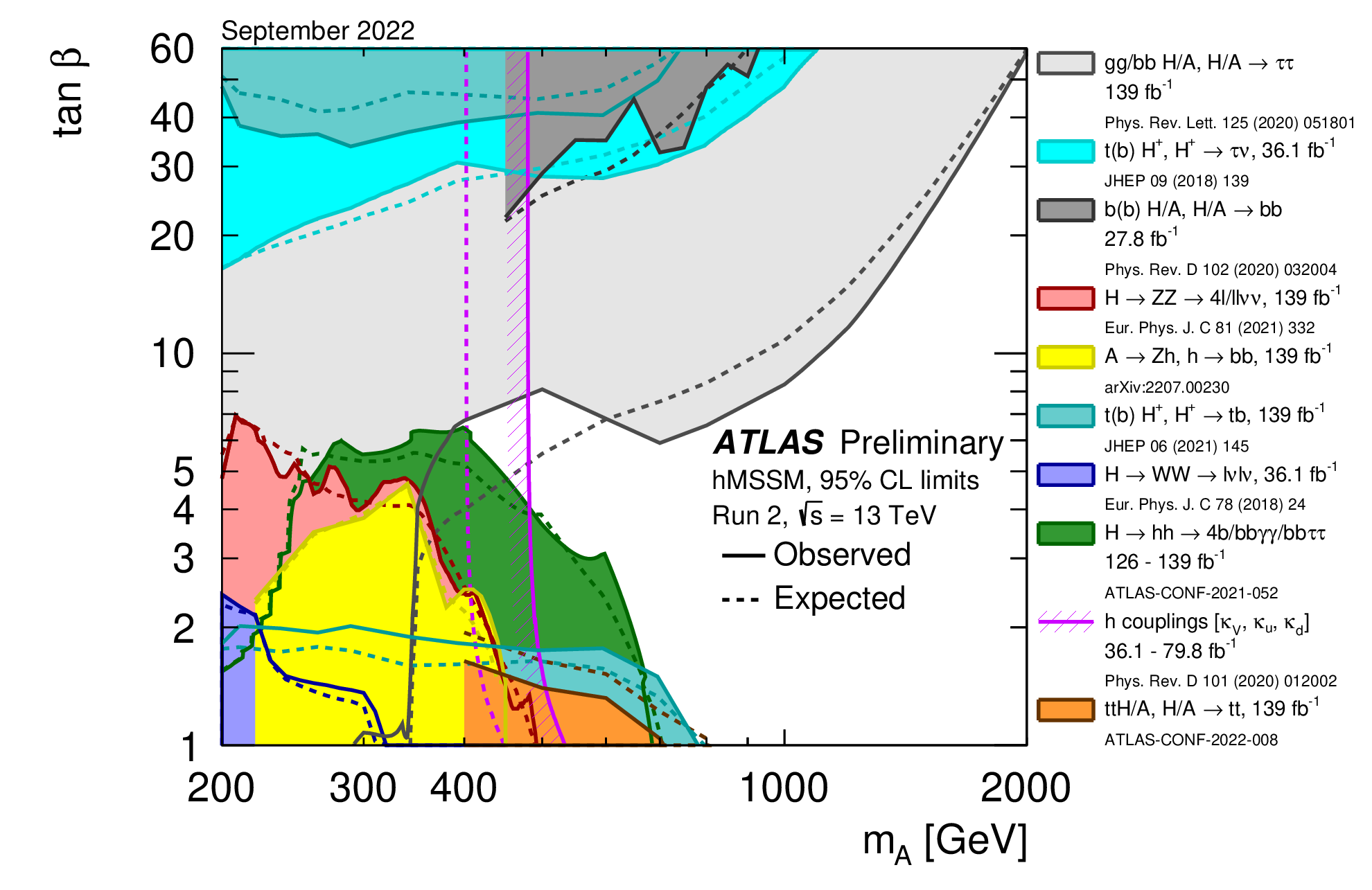}  }
\vspace*{-.3cm}
\caption[]{Regions of the $[M_A, \tan\beta]$ plane in a benchmark scenario of the minimal supersymmetric extension of the SM  called hMSSM \cite{Djouadi:2013uqa}  that are  excluded via direct searches for the heavy $A,H,H^\pm$ bosons in various channels; the observed and expected limits are in solid and dashed lines. Also shown are the constraints from the fits to the measured rates of the observed SM-like $h$ state in all production and decay modes;  they are quoted at 95\% confidence. The full data collected in Run II at an energy of 13 TeV is used. From Ref.~\cite{ATLAS:2022vkf}.} 
\label{fig:hMSSM}
\vspace*{-.2cm}
\end{figure}

All these extensions were expected to lead to a rich spectrum  of new particles with masses that are only slightly above the Fermi scale and, hence, within the reach of the  LHC. Alas, no new state of any type has been discovered and, at present, there is no experimental measurement that deviates from its predicted value and the SM is standing chiefly triumphant and unaltered. As an example, we display in Fig.~\ref{fig:hMSSM}, the constraints that are imposed by LHC searches on the additional Higgs particles of the minimal supersymmetric SM which is a Type-II 2HDM that can be described (in a scenario called hMSSM \cite{Djouadi:2013uqa}) by two basic parameters introduced earlier: the ratio of vevs $\tan\beta$ and the pseudoscalar $A$ boson mass, $M_A\! \approx\! M_H\! \approx\! M_{H^\pm}$. As can be seen, masses of the extra $A,H,H^\pm$ bosons well above the TeV scale are now excluded for some values of the parameter $\tan\beta$. 

A related problem to this absence of a new physics signal is that, for an increasing number of people, it looks rather unlikely that  the slight upgrade in center of mass energy and the order of magnitude increase of luminosity that are planned for the next campaign of searches at the LHC, the so-called high-luminosity run,  would allow for the observation of a new phenomenon in a very conclusive manner. There is, nevertheless, at least one aspect which is largely untested and which could provide us with some surprise: the Higgs self-coupling which can be probed only in double Higgs production which has rather low rates \cite{Baglio:2012np}. The determination of this coupling is at the top of the agenda of the high-luminosity option. Despite of this, there is a high probability that we will have to await for  the next generation of high-energy experiments, either a lepton collider with a c.m. energy from the Fermi to the TeV scale, or a proton collider with an energy up to 100 TeV, to witness a new discovery. Unfortunately, these new  machines, if eventually financed, are expected to be made available and to operate only in a few decades from now.  This poses a serious threat to the whole field of high-energy physics. 

All what we can do in the meantime is to test and retest the SM and to search indirectly for any type of new physics in the most precise, most general and most model-independent way. 

For this purpose, we have an extremely powerful tool that Weinberg put at our disposal again in the 1970s: the effective field theory or EFT  \cite{Weinberg:1979sa,Weinberg:1978kz}.  In this approach, physical processes at current energies are described by an ${\SU3_{c} \times \SU2_{L} \times \U1_{Y}}$ invariant effective theory obtained by integrating out all the heavy degrees of freedom, with masses at the characteristic scale $\Lambda$, and involving only the light SM particles. This is just in the same way as the old point-like four fermion Fermi theory, which is not renormalizable nor unitary at high energies, is the EFT of the weak interactions at energies much below the Fermi scale, $\Lambda\!=\!v \!\approx \!250$ GeV. The Lagrangian contains the SM one, but also an infinite set of non-renormalizable higher-dimensional gauge invariant operators organized in a systematic expansion in terms of the heavy scale $\Lambda^{D-4}$ with $D$ the canonical dimension of the operator. In practice, the leading contributions beyond the SM ones are those of the operators of dimension $5$ and $6$ and all higher dimension operators should be irrelevant.
This framework provides a nice way of splitting in two steps the comparison between experiment and theory and obtain information on possible SM extensions. In a first one, experimental observables are encoded in terms of the effective operator coefficients with a minimal theoretical bias. In a second step, the coefficients of the operators are connected to the parameters of specific new physics models through a process called matching.
This SMEFT is now becoming the privileged tool to test the SM and to search for new physics beyond it; see e.g. Ref.~\cite{deBlas:2017xtg} for a nice review. 

As for the solution of the notorious hierarchy problem, which was the driving criterium for selecting new physics theories in the last four decades,  one might momentarily abandon invoking it and simply resort to the same argument that Weinberg put forward to explain the much worse problem of the smallness of the cosmological constant \cite{Weinberg:1987dv}: the anthropic principle. 

From the preceding discussion, it should by now become clear that there is an extraordinary large number of key aspects  of modern particle physics that were shaped by ideas and sharpened by methods that Steven Weinberg put forward. For this reason, he should clearly be considered as one of the greatest figures of our field and probably the most prominent one in the period between the early 1960s to the early 1980s, which witnessed some of the most important achievements in particle physics, particularly from the theoretical side. This is indubitably the case concerning the field of Higgs physics in which he left an unmistakable and profound mark.\clearpage 

\noindent {\bf Acknowledgements:} AD is supported by the Junta de Andalucia through the Talentia Senior program and the grant PID2021-128396NB-I00. JII is supported by the Spanish Ministry of Science, Innovation and Universities under the grant PID2022-140440NB-C21. 

\bibliographystyle{elsarticle-num}
\bibliography{references}

\begin{thebibliography}{10}
\expandafter\ifx\csname url\endcsname\relax
  \def\url#1{\texttt{#1}}\fi
\expandafter\ifx\csname urlprefix\endcsname\relax\def\urlprefix{URL }\fi
\expandafter\ifx\csname href\endcsname\relax
  \def\href#1#2{#2} \def\path#1{#1}\fi

\bibitem{Goldstone:1962es}
J.~Goldstone, A.~Salam, S.~Weinberg, {Broken Symmetries}, Phys. Rev. 127 (1962)
  965--970.
\newblock \href {https://doi.org/10.1103/PhysRev.127.965}
  {\path{doi:10.1103/PhysRev.127.965}}.

\bibitem{Weinberg:1967tq}
S.~Weinberg, {A Model of Leptons}, Phys. Rev. Lett. 19 (1967) 1264--1266.
\newblock \href {https://doi.org/10.1103/PhysRevLett.19.1264}
  {\path{doi:10.1103/PhysRevLett.19.1264}}.

\bibitem{Weinberg:1976pe}
S.~Weinberg, {Mass of the Higgs Boson}, Phys. Rev. Lett. 36 (1976) 294--296.
\newblock \href {https://doi.org/10.1103/PhysRevLett.36.294}
  {\path{doi:10.1103/PhysRevLett.36.294}}.

\bibitem{Gildener:1976ih}
E.~Gildener, S.~Weinberg, {Symmetry Breaking and Scalar Bosons}, Phys. Rev. D
  13 (1976) 3333.
\newblock \href {https://doi.org/10.1103/PhysRevD.13.3333}
  {\path{doi:10.1103/PhysRevD.13.3333}}.

\bibitem{Glashow:1976nt}
S.~L. Glashow, S.~Weinberg, {Natural Conservation Laws for Neutral Currents},
  Phys. Rev. D 15 (1977) 1958.
\newblock \href {https://doi.org/10.1103/PhysRevD.15.1958}
  {\path{doi:10.1103/PhysRevD.15.1958}}.

\bibitem{Weinberg:1976hu}
S.~Weinberg, {Gauge Theory of CP Violation}, Phys. Rev. Lett. 37 (1976) 657.
\newblock \href {https://doi.org/10.1103/PhysRevLett.37.657}
  {\path{doi:10.1103/PhysRevLett.37.657}}.

\bibitem{Weinberg:1989dx}
S.~Weinberg, {Larger Higgs Exchange Terms in the Neutron Electric Dipole
  Moment}, Phys. Rev. Lett. 63 (1989) 2333.
\newblock \href {https://doi.org/10.1103/PhysRevLett.63.2333}
  {\path{doi:10.1103/PhysRevLett.63.2333}}.

\bibitem{Weinberg:1990me}
S.~Weinberg, {Unitarity Constraints on {CP} Nonconservation in Higgs Exchange},
  Phys. Rev. D 42 (1990) 860--866.
\newblock \href {https://doi.org/10.1103/PhysRevD.42.860}
  {\path{doi:10.1103/PhysRevD.42.860}}.

\bibitem{Aad:2012tfa}
G.~Aad, et~al., {Observation of a new particle in the search for the Standard
  Model Higgs boson with the ATLAS detector at the LHC}, Phys. Lett. B716
  (2012) 1--29.
\newblock \href {http://arxiv.org/abs/1207.7214} {\path{arXiv:1207.7214}},
  \href {https://doi.org/10.1016/j.physletb.2012.08.020}
  {\path{doi:10.1016/j.physletb.2012.08.020}}.

\bibitem{Chatrchyan:2012xdj}
S.~Chatrchyan, et~al., {Observation of a New Boson at a Mass of 125 GeV with
  the CMS Experiment at the LHC}, Phys. Lett. B716 (2012) 30--61.
\newblock \href {http://arxiv.org/abs/1207.7235} {\path{arXiv:1207.7235}},
  \href {https://doi.org/10.1016/j.physletb.2012.08.021}
  {\path{doi:10.1016/j.physletb.2012.08.021}}.

\bibitem{ATLAS:2022vkf}
G.~Aad, et~al., {A detailed map of Higgs boson interactions by the ATLAS
  experiment ten years after the discovery}, Nature 607~(7917) (2022) 52--59,
  [Erratum: Nature 612, E24 (2022)].
\newblock \href {http://arxiv.org/abs/2207.00092} {\path{arXiv:2207.00092}},
  \href {https://doi.org/10.1038/s41586-022-04893-w}
  {\path{doi:10.1038/s41586-022-04893-w}}.

\bibitem{CMS:2022dwd}
A.~Tumasyan, et~al., {A portrait of the Higgs boson by the CMS experiment ten
  years after the discovery.}, Nature 607~(7917) (2022) 60--68.
\newblock \href {http://arxiv.org/abs/2207.00043} {\path{arXiv:2207.00043}},
  \href {https://doi.org/10.1038/s41586-022-04892-x}
  {\path{doi:10.1038/s41586-022-04892-x}}.

\bibitem{Weinberg:2018apv}
S.~Weinberg, {Essay: Half a Century of the Standard Model}, Phys. Rev. Lett.
  121~(22) (2018) 220001.
\newblock \href {https://doi.org/10.1103/PhysRevLett.121.220001}
  {\path{doi:10.1103/PhysRevLett.121.220001}}.

\bibitem{Weinberg:1975gm}
S.~Weinberg, {Implications of Dynamical Symmetry Breaking}, Phys. Rev. D 13
  (1976) 974--996, [Add: Phys.Rev.D 19, 1277--1280 (1979)].
\newblock \href {https://doi.org/10.1103/PhysRevD.19.1277}
  {\path{doi:10.1103/PhysRevD.19.1277}}.

\bibitem{Weinberg:1981wj}
S.~Weinberg, {Supersymmetry at Ordinary Energies. 1. Masses and Conservation
  Laws}, Phys. Rev. D 26 (1982) 287.
\newblock \href {https://doi.org/10.1103/PhysRevD.26.287}
  {\path{doi:10.1103/PhysRevD.26.287}}.

\bibitem{Farrar:1982te}
G.~R. Farrar, S.~Weinberg, {Supersymmetry at Ordinary Energies. 2. R
  Invariance, Golsdtone ....}, Phys. Rev. D 27 (1983) 2732.
\newblock \href {https://doi.org/10.1103/PhysRevD.27.2732}
  {\path{doi:10.1103/PhysRevD.27.2732}}.

\bibitem{Hall:1983iz}
L.~Hall, J.~Lykken, S.~Weinberg, {Supergravity as the Messenger of
  Supersymmetry Breaking}, Phys. Rev. D 27 (1983) 2359--2378.
\newblock \href {https://doi.org/10.1103/PhysRevD.27.2359}
  {\path{doi:10.1103/PhysRevD.27.2359}}.

\bibitem{Gell-Mann:1961omu}
M.~Gell-Mann, {The Eightfold Way: A Theory of strong interaction symmetry} (3
  1961).
\newblock \href {https://doi.org/10.2172/4008239} {\path{doi:10.2172/4008239}}.

\bibitem{Nambu:1960tm}
Y.~Nambu, {Quasiparticles and Gauge Invariance in the Theory of
  Superconductivity}, Phys. Rev. 117 (1960) 648--663.
\newblock \href {https://doi.org/10.1103/PhysRev.117.648}
  {\path{doi:10.1103/PhysRev.117.648}}.

\bibitem{Goldstone:1961eq}
J.~Goldstone, {Field Theories with Superconductor Solutions}, Nuovo Cim. 19
  (1961) 154--164.
\newblock \href {https://doi.org/10.1007/BF02812722}
  {\path{doi:10.1007/BF02812722}}.

\bibitem{Higgs:1964ia}
P.~W. Higgs, {Broken symmetries, massless particles and gauge fields}, Phys.
  Lett. 12 (1964) 132--133.
\newblock \href {https://doi.org/10.1016/0031-9163(64)91136-9}
  {\path{doi:10.1016/0031-9163(64)91136-9}}.

\bibitem{Englert:1964et}
F.~Englert, R.~Brout, {Broken Symmetry and the Mass of Gauge Vector Mesons},
  Phys. Rev. Lett. 13 (1964) 321--323.
\newblock \href {https://doi.org/10.1103/PhysRevLett.13.321}
  {\path{doi:10.1103/PhysRevLett.13.321}}.

\bibitem{Guralnik:1964eu}
G.~S. Guralnik, C.~R. Hagen, T.~W.~B. Kibble, {Global Conservation Laws and
  Massless Particles}, Phys. Rev. Lett. 13 (1964) 585--587.
\newblock \href {https://doi.org/10.1103/PhysRevLett.13.585}
  {\path{doi:10.1103/PhysRevLett.13.585}}.

\bibitem{Higgs:1964pj}
P.~W. Higgs, {Broken Symmetries and the Masses of Gauge Bosons}, Phys. Rev.
  Lett. 13 (1964) 508--509.
\newblock \href {https://doi.org/10.1103/PhysRevLett.13.508}
  {\path{doi:10.1103/PhysRevLett.13.508}}.

\bibitem{Higgs:1966ev}
P.~W. Higgs, {Spontaneous Symmetry Breakdown without Massless Bosons}, Phys.
  Rev. 145 (1966) 1156--1163.
\newblock \href {https://doi.org/10.1103/PhysRev.145.1156}
  {\path{doi:10.1103/PhysRev.145.1156}}.

\bibitem{Fermi:1934hr}
E.~Fermi, {An attempt of a theory of beta radiation. 1.}, Z. Phys. 88 (1934)
  161--177.
\newblock \href {https://doi.org/10.1007/BF01351864}
  {\path{doi:10.1007/BF01351864}}.

\bibitem{Feynman:1958ty}
R.~P. Feynman, M.~Gell-Mann, {Theory of Fermi interaction}, Phys. Rev. 109
  (1958) 193--198.
\newblock \href {https://doi.org/10.1103/PhysRev.109.193}
  {\path{doi:10.1103/PhysRev.109.193}}.

\bibitem{Sudarshan:1958vf}
E.~C.~G. Sudarshan, R.~e. Marshak, {Chirality invariance and the universal
  Fermi interaction}, Phys. Rev. 109 (1958) 1860--1860.
\newblock \href {https://doi.org/10.1103/PhysRev.109.1860.2}
  {\path{doi:10.1103/PhysRev.109.1860.2}}.

\bibitem{Yang:1954ek}
C.-N. Yang, R.~L. Mills, {Conservation of Isotopic Spin and Isotopic Gauge
  Invariance}, Phys. Rev. 96 (1954) 191--195.
\newblock \href {https://doi.org/10.1103/PhysRev.96.191}
  {\path{doi:10.1103/PhysRev.96.191}}.

\bibitem{Jackson:2001ia}
J.~D. Jackson, L.~B. Okun, {Historical roots of gauge invariance}, Rev. Mod.
  Phys. 73 (2001) 663--680.
\newblock \href {http://arxiv.org/abs/hep-ph/0012061}
  {\path{arXiv:hep-ph/0012061}}, \href
  {https://doi.org/10.1103/RevModPhys.73.663}
  {\path{doi:10.1103/RevModPhys.73.663}}.

\bibitem{Glashow:1961tr}
S.~L. Glashow, {Partial Symmetries of Weak Interactions}, Nucl. Phys. 22 (1961)
  579--588.
\newblock \href {https://doi.org/10.1016/0029-5582(61)90469-2}
  {\path{doi:10.1016/0029-5582(61)90469-2}}.

\bibitem{Salam:1964ry}
A.~Salam, J.~C. Ward, {Electromagnetic and weak interactions}, Phys. Lett. 13
  (1964) 168--171.
\newblock \href {https://doi.org/10.1016/0031-9163(64)90711-5}
  {\path{doi:10.1016/0031-9163(64)90711-5}}.

\bibitem{Schwinger:1957em}
J.~S. Schwinger, {A Theory of the Fundamental Interactions}, Annals Phys. 2
  (1957) 407--434.
\newblock \href {https://doi.org/10.1016/0003-4916(57)90015-5}
  {\path{doi:10.1016/0003-4916(57)90015-5}}.

\bibitem{Salam:1968rm}
A.~Salam, {Weak and Electromagnetic Interactions}, Conf. Proc. C 680519 (1968)
  367--377.
\newblock \href {https://doi.org/10.1142/9789812795915-0034}
  {\path{doi:10.1142/9789812795915-0034}}.

\bibitem{Illana:2022hab}
J.~I. Illana, A.~Jimenez~Cano, {Quantum field theory and the structure of the
  Standard Model}, PoS CORFU2021 (2022) 314.
\newblock \href {http://arxiv.org/abs/2211.14636} {\path{arXiv:2211.14636}},
  \href {https://doi.org/10.22323/1.406.0314} {\path{doi:10.22323/1.406.0314}}.

\bibitem{Zweig:1964ruk}
G.~Zweig, {An SU(3) model for strong interaction symmetry and its breaking.
  Version 1} (1 1964).

\bibitem{Bloom:1969kc}
E.~D. Bloom, et~al., {High-Energy Inelastic e p Scattering at 6-Degrees and
  10-Degrees}, Phys. Rev. Lett. 23 (1969) 930--934.
\newblock \href {https://doi.org/10.1103/PhysRevLett.23.930}
  {\path{doi:10.1103/PhysRevLett.23.930}}.

\bibitem{Breidenbach:1969kd}
M.~Breidenbach, J.~I. Friedman, H.~W. Kendall, E.~D. Bloom, D.~H. Coward, H.~C.
  DeStaebler, J.~Drees, L.~W. Mo, R.~E. Taylor, {Observed behavior of highly
  inelastic electron-proton scattering}, Phys. Rev. Lett. 23 (1969) 935--939.
\newblock \href {https://doi.org/10.1103/PhysRevLett.23.935}
  {\path{doi:10.1103/PhysRevLett.23.935}}.

\bibitem{E598:1974sol}
J.~J. Aubert, et~al., {Experimental Observation of a Heavy Particle $J$}, Phys.
  Rev. Lett. 33 (1974) 1404--1406.
\newblock \href {https://doi.org/10.1103/PhysRevLett.33.1404}
  {\path{doi:10.1103/PhysRevLett.33.1404}}.

\bibitem{SLAC-SP-017:1974ind}
J.~E. Augustin, et~al., {Discovery of a Narrow Resonance in $e^+ e^-$
  Annihilation}, Phys. Rev. Lett. 33 (1974) 1406--1408.
\newblock \href {https://doi.org/10.1103/PhysRevLett.33.1406}
  {\path{doi:10.1103/PhysRevLett.33.1406}}.

\bibitem{Perl:1975bf}
M.~L. Perl, et~al., {Evidence for Anomalous Lepton Production in e+ - e-
  Annihilation}, Phys. Rev. Lett. 35 (1975) 1489--1492.
\newblock \href {https://doi.org/10.1103/PhysRevLett.35.1489}
  {\path{doi:10.1103/PhysRevLett.35.1489}}.

\bibitem{E288:1977xhf}
S.~W. Herb, et~al., {Observation of a Dimuon Resonance at 9.5-GeV in 400-GeV
  Proton-Nucleus Collisions}, Phys. Rev. Lett. 39 (1977) 252--255.
\newblock \href {https://doi.org/10.1103/PhysRevLett.39.252}
  {\path{doi:10.1103/PhysRevLett.39.252}}.

\bibitem{CDF:1995wbb}
F.~Abe, et~al., {Observation of top quark production in $\bar{p}p$ collisions},
  Phys. Rev. Lett. 74 (1995) 2626--2631.
\newblock \href {http://arxiv.org/abs/hep-ex/9503002}
  {\path{arXiv:hep-ex/9503002}}, \href
  {https://doi.org/10.1103/PhysRevLett.74.2626}
  {\path{doi:10.1103/PhysRevLett.74.2626}}.

\bibitem{D0:1995jca}
S.~Abachi, et~al., {Observation of the top quark}, Phys. Rev. Lett. 74 (1995)
  2632--2637.
\newblock \href {http://arxiv.org/abs/hep-ex/9503003}
  {\path{arXiv:hep-ex/9503003}}, \href
  {https://doi.org/10.1103/PhysRevLett.74.2632}
  {\path{doi:10.1103/PhysRevLett.74.2632}}.

\bibitem{Cabibbo:1963yz}
N.~Cabibbo, {Unitary Symmetry and Leptonic Decays}, Phys. Rev. Lett. 10 (1963)
  531--533.
\newblock \href {https://doi.org/10.1103/PhysRevLett.10.531}
  {\path{doi:10.1103/PhysRevLett.10.531}}.

\bibitem{Kobayashi:1973fv}
M.~Kobayashi, T.~Maskawa, {CP Violation in the renormalizable theory of weak
  interaction}, Prog. Theor. Phys. 49 (1973) 652--657.
\newblock \href {https://doi.org/10.1143/PTP.49.652}
  {\path{doi:10.1143/PTP.49.652}}.

\bibitem{Glashow:1970gm}
S.~L. Glashow, J.~Iliopoulos, L.~Maiani, {Weak Interactions with Lepton-Hadron
  Symmetry}, Phys. Rev. D 2 (1970) 1285--1292.
\newblock \href {https://doi.org/10.1103/PhysRevD.2.1285}
  {\path{doi:10.1103/PhysRevD.2.1285}}.

\bibitem{Faddeev:1967fc}
L.~Faddeev, V.~Popov, {Feynman Diagrams for the Yang-Mills Field}, Phys. Lett.
  B 25 (1967) 29--30.
\newblock \href {https://doi.org/10.1016/0370-2693(67)90067-6}
  {\path{doi:10.1016/0370-2693(67)90067-6}}.

\bibitem{tHooft:1971qjg}
G.~'t~Hooft, {Renormalizable Lagrangians for Massive Yang-Mills Fields}, Nucl.
  Phys. B 35 (1971) 167.
\newblock \href {https://doi.org/10.1016/0550-3213(71)90139-8}
  {\path{doi:10.1016/0550-3213(71)90139-8}}.

\bibitem{tHooft:1972tcz}
G.~'t~Hooft, M.~J.~G. Veltman, {Regularization and Renormalization of Gauge
  Fields}, Nucl. Phys. B 44 (1972) 189--213.
\newblock \href {https://doi.org/10.1016/0550-3213(72)90279-9}
  {\path{doi:10.1016/0550-3213(72)90279-9}}.

\bibitem{Lee:1972ocr}
B.~W. Lee, J.~Zinn-Justin, {Spontaneously Broken Gauge Symmetries Part 2:
  Perturbation Theory and Renormalization}, Phys. Rev. D 5 (1972) 3137--3155,
  [Erratum: Phys.Rev.D 8, 4654 (1973)].
\newblock \href {https://doi.org/10.1103/PhysRevD.5.3137}
  {\path{doi:10.1103/PhysRevD.5.3137}}.

\bibitem{Coleman:1979}
S.~R. Coleman, {The 1979 Nobel Prize in Physics}, in: {Science, New Series,
  Vol. 206, No. 4424}, 1979, pp. 1290--1292.

\bibitem{Abers:1973qs}
E.~S. Abers, B.~W. Lee, {Gauge Theories}, Phys. Rept. 9 (1973) 1--141.
\newblock \href {https://doi.org/10.1016/0370-1573(73)90027-6}
  {\path{doi:10.1016/0370-1573(73)90027-6}}.

\bibitem{GargamelleNeutrino:1973jyy}
F.~J. Hasert, et~al., {Observation of Neutrino Like Interactions Without Muon
  Or Electron in the Gargamelle Neutrino Experiment}, Phys. Lett. B 46 (1973)
  138--140.
\newblock \href {https://doi.org/10.1016/0370-2693(73)90499-1}
  {\path{doi:10.1016/0370-2693(73)90499-1}}.

\bibitem{UA1:1983crd}
G.~Arnison, et~al., {Experimental Observation of Isolated Large Transverse
  Energy Electrons with Associated Missing Energy at $\sqrt{s}= 540$ GeV},
  Phys. Lett. B 122 (1983) 103--116.
\newblock \href {https://doi.org/10.1016/0370-2693(83)91177-2}
  {\path{doi:10.1016/0370-2693(83)91177-2}}.

\bibitem{UA2:1983tsx}
M.~Banner, et~al., {Observation of Single Isolated Electrons of High Transverse
  Momentum in Events with Missing Transverse Energy at the CERN anti-p p
  Collider}, Phys. Lett. B 122 (1983) 476--485.
\newblock \href {https://doi.org/10.1016/0370-2693(83)91605-2}
  {\path{doi:10.1016/0370-2693(83)91605-2}}.

\bibitem{UA1:1983mne}
G.~Arnison, et~al., {Experimental Observation of Lepton Pairs of Invariant Mass
  Around 95-GeV at the CERN SPS Collider}, Phys. Lett. B 126 (1983) 398--410.
\newblock \href {https://doi.org/10.1016/0370-2693(83)90188-0}
  {\path{doi:10.1016/0370-2693(83)90188-0}}.

\bibitem{UA2:1983mlz}
P.~Bagnaia, et~al., {Evidence for $Z^{0} \to e^+ e^-$ at the CERN $\bar{p} p$
  Collider}, Phys.Lett.B 129 (1983) 130.
\newblock \href {https://doi.org/10.1016/0370-2693(83)90744-X}
  {\path{doi:10.1016/0370-2693(83)90744-X}}.

\bibitem{Linde:1975sw}
A.~Linde, {Dynamical Symmetry Restoration and Constraints on Masses and
  Coupling Constants in Gauge Theories}, JETP Lett. 23 (1976) 64--67.

\bibitem{Cabibbo:1979ay}
N.~Cabibbo, L.~Maiani, G.~Parisi, R.~Petronzio, {Bounds on the Fermions and
  Higgs Boson Masses in Grand Unified Theories}, Nucl. Phys. B 158 (1979)
  295--305.
\newblock \href {https://doi.org/10.1016/0550-3213(79)90167-6}
  {\path{doi:10.1016/0550-3213(79)90167-6}}.

\bibitem{Politzer:1978ic}
H.~D. Politzer, S.~Wolfram, {Bounds on Particle Masses in the Weinberg-Salam
  Model}, Phys. Lett. B 82 (1979) 242--246, [Erratum: Phys.Lett.B 83, 421
  (1979)].
\newblock \href {https://doi.org/10.1016/0370-2693(79)90746-9}
  {\path{doi:10.1016/0370-2693(79)90746-9}}.

\bibitem{Hung:1979dn}
P.~Q. Hung, {Vacuum Instability and New Constraints on Fermion Masses}, Phys.
  Rev. Lett. 42 (1979) 873.
\newblock \href {https://doi.org/10.1103/PhysRevLett.42.873}
  {\path{doi:10.1103/PhysRevLett.42.873}}.

\bibitem{Sher:1988mj}
M.~Sher, {Electroweak Higgs Potentials and Vacuum Stability}, Phys. Rept. 179
  (1989) 273--418.
\newblock \href {https://doi.org/10.1016/0370-1573(89)90061-6}
  {\path{doi:10.1016/0370-1573(89)90061-6}}.

\bibitem{Dicus:1973gbw}
D.~A. Dicus, V.~S. Mathur, {Upper bounds on the values of masses in unified
  gauge theories}, Phys. Rev. D 7 (1973) 3111--3114.
\newblock \href {https://doi.org/10.1103/PhysRevD.7.3111}
  {\path{doi:10.1103/PhysRevD.7.3111}}.

\bibitem{Cornwall:1974km}
J.~M. Cornwall, D.~N. Levin, G.~Tiktopoulos, {Derivation of Gauge Invariance
  from High-Energy Unitarity Bounds on the s Matrix}, Phys. Rev. D 10 (1974)
  1145, [Erratum: Phys.Rev.D 11, 972 (1975)].
\newblock \href {https://doi.org/10.1103/PhysRevD.10.1145}
  {\path{doi:10.1103/PhysRevD.10.1145}}.

\bibitem{Veltman:1976rt}
M.~J.~G. Veltman, {Second Threshold in Weak Interactions}, Acta Phys. Polon. B
  8 (1977) 475--492.

\bibitem{Lee:1977eg}
B.~Lee, C.~Quigg, H.~Thacker, {Weak interactions at very high-energies: the
  role of the Higgs boson mass}, Phys. Rev. D 16 (1977) 1519.
\newblock \href {https://doi.org/10.1103/PhysRevD.16.1519}
  {\path{doi:10.1103/PhysRevD.16.1519}}.

\bibitem{Wilson:1973jj}
K.~G. Wilson, J.~B. Kogut, {The Renormalization group and the epsilon
  expansion}, Phys. Rept. 12 (1974) 75--199.
\newblock \href {https://doi.org/10.1016/0370-1573(74)90023-4}
  {\path{doi:10.1016/0370-1573(74)90023-4}}.

\bibitem{Hambye:1996wb}
T.~Hambye, K.~Riesselmann, {Matching conditions and Higgs mass upper bounds
  revisited}, Phys. Rev. D 55 (1997) 7255--7262.
\newblock \href {http://arxiv.org/abs/hep-ph/9610272}
  {\path{arXiv:hep-ph/9610272}}, \href
  {https://doi.org/10.1103/PhysRevD.55.7255}
  {\path{doi:10.1103/PhysRevD.55.7255}}.

\bibitem{Degrassi:2012ry}
G.~Degrassi, S.~Di~Vita, J.~Elias-Miro, J.~Espinosa, G.~Giudice, G.~Isidori,
  A.~Strumia, {Higgs mass and vacuum stability in the Standard Model at NNLO},
  JHEP 08 (2012) 098.
\newblock \href {http://arxiv.org/abs/1205.6497} {\path{arXiv:1205.6497}},
  \href {https://doi.org/10.1007/JHEP08(2012)098}
  {\path{doi:10.1007/JHEP08(2012)098}}.

\bibitem{Bezrukov:2012sa}
F.~Bezrukov, M.~Kalmykov, B.~Kniehl, M.~Shaposhnikov, {Higgs Boson Mass and New
  Physics}, JHEP 10 (2012) 140.
\newblock \href {http://arxiv.org/abs/1205.2893} {\path{arXiv:1205.2893}},
  \href {https://doi.org/10.1007/JHEP10(2012)140}
  {\path{doi:10.1007/JHEP10(2012)140}}.

\bibitem{Alekhin:2012py}
S.~Alekhin, A.~Djouadi, S.~Moch, {The top quark and Higgs boson masses and
  stability of the electroweak vacuum}, Phys. Lett. B 716 (2012) 214--219.
\newblock \href {http://arxiv.org/abs/1207.0980} {\path{arXiv:1207.0980}},
  \href {https://doi.org/10.1016/j.physletb.2012.08.024}
  {\path{doi:10.1016/j.physletb.2012.08.024}}.

\bibitem{Veltman:1977kh}
M.~Veltman, {Limit on Mass Differences in the Weinberg Model}, Nucl. Phys. B
  123 (1977) 89--99.
\newblock \href {https://doi.org/10.1016/0550-3213(77)90342-X}
  {\path{doi:10.1016/0550-3213(77)90342-X}}.

\bibitem{Paschos:1976ay}
E.~A. Paschos, {Diagonal Neutral Currents}, Phys. Rev. D 15 (1977) 1966.
\newblock \href {https://doi.org/10.1103/PhysRevD.15.1966}
  {\path{doi:10.1103/PhysRevD.15.1966}}.

\bibitem{Aoki:2009ha}
M.~Aoki, S.~Kanemura, K.~Tsumura, K.~Yagyu, {Models of Yukawa interaction in
  the two Higgs doublet model, and their collider phenomenology}, Phys. Rev. D
  80 (2009) 015017.
\newblock \href {http://arxiv.org/abs/0902.4665} {\path{arXiv:0902.4665}},
  \href {https://doi.org/10.1103/PhysRevD.80.015017}
  {\path{doi:10.1103/PhysRevD.80.015017}}.

\bibitem{Branco:2011iw}
G.~Branco, P.~Ferreira, L.~Lavoura, M.~Rebelo, M.~Sher, J.~Silva, {Theory and
  phenomenology of two-Higgs-doublet models}, Phys. Rept. 516 (2012) 1--102.
\newblock \href {http://arxiv.org/abs/1106.0034} {\path{arXiv:1106.0034}},
  \href {https://doi.org/10.1016/j.physrep.2012.02.002}
  {\path{doi:10.1016/j.physrep.2012.02.002}}.

\bibitem{Gunion:1989we}
J.~F. Gunion, H.~E. Haber, G.~L. Kane, S.~Dawson, {The Higgs hunter's guide},
  Perseus Publishing, 1990.

\bibitem{Djouadi:2005gj}
A.~Djouadi, {The Anatomy of electro-weak symmetry breaking. II. The Higgs
  bosons in the minimal supersymmetric model}, Phys. Rept. 459 (2008) 1--241.
\newblock \href {http://arxiv.org/abs/hep-ph/0503173}
  {\path{arXiv:hep-ph/0503173}}, \href
  {https://doi.org/10.1016/j.physrep.2007.10.005}
  {\path{doi:10.1016/j.physrep.2007.10.005}}.

\bibitem{Arcadi:2019lka}
G.~Arcadi, A.~Djouadi, M.~Raidal, {Dark Matter through the Higgs portal}, Phys.
  Rept. 842 (2020) 1--180.
\newblock \href {http://arxiv.org/abs/1903.03616} {\path{arXiv:1903.03616}},
  \href {https://doi.org/10.1016/j.physrep.2019.11.003}
  {\path{doi:10.1016/j.physrep.2019.11.003}}.

\bibitem{Gross:1973id}
D.~J. Gross, F.~Wilczek, {Ultraviolet Behavior of Nonabelian Gauge Theories},
  Phys. Rev. Lett. 30 (1973) 1343--1346.
\newblock \href {https://doi.org/10.1103/PhysRevLett.30.1343}
  {\path{doi:10.1103/PhysRevLett.30.1343}}.

\bibitem{Politzer:1973fx}
H.~D. Politzer, {Reliable Perturbative Results for Strong Interactions?}, Phys.
  Rev. Lett. 30 (1973) 1346--1349.
\newblock \href {https://doi.org/10.1103/PhysRevLett.30.1346}
  {\path{doi:10.1103/PhysRevLett.30.1346}}.

\bibitem{TASSO:1979zyf}
R.~Brandelik, et~al., {Evidence for Planar Events in e+ e- Annihilation at
  High-Energies}, Phys. Lett. B 86 (1979) 243--249.
\newblock \href {https://doi.org/10.1016/0370-2693(79)90830-X}
  {\path{doi:10.1016/0370-2693(79)90830-X}}.

\bibitem{ParticleDataGroup:2012pjm}
J.~Beringer, et~al., {Review of Particle Physics (RPP)}, Phys. Rev. D 86 (2012)
  010001.
\newblock \href {https://doi.org/10.1103/PhysRevD.86.010001}
  {\path{doi:10.1103/PhysRevD.86.010001}}.

\bibitem{Ellis:1975ap}
J.~R. Ellis, M.~K. Gaillard, D.~V. Nanopoulos, {A Phenomenological Profile of
  the Higgs Boson}, Nucl. Phys. B 106 (1976) 292.
\newblock \href {https://doi.org/10.1016/0550-3213(76)90382-5}
  {\path{doi:10.1016/0550-3213(76)90382-5}}.

\bibitem{Djouadi:2005gi}
A.~Djouadi, {The Anatomy of electro-weak symmetry breaking. I: The Higgs boson
  in the standard model}, Phys. Rept. 457 (2008) 1--216.
\newblock \href {http://arxiv.org/abs/hep-ph/0503172}
  {\path{arXiv:hep-ph/0503172}}, \href
  {https://doi.org/10.1016/j.physrep.2007.10.004}
  {\path{doi:10.1016/j.physrep.2007.10.004}}.

\bibitem{LHCHiggsCrossSectionWorkingGroup:2011wcg}
S.~Dittmaier, et~al., {Handbook of LHC Higgs Cross Sections: 1. Inclusive
  Observables} (1 2011).
\newblock \href {http://arxiv.org/abs/1101.0593} {\path{arXiv:1101.0593}},
  \href {https://doi.org/10.5170/CERN-2011-002}
  {\path{doi:10.5170/CERN-2011-002}}.

\bibitem{Weinberg:2020zba}
S.~Weinberg, {Models of Lepton and Quark Masses}, Phys. Rev. D 101~(3) (2020)
  035020.
\newblock \href {http://arxiv.org/abs/2001.06582} {\path{arXiv:2001.06582}},
  \href {https://doi.org/10.1103/PhysRevD.101.035020}
  {\path{doi:10.1103/PhysRevD.101.035020}}.

\bibitem{Weinberg:1979sa}
S.~Weinberg, {Baryon and Lepton Nonconserving Processes}, Phys. Rev. Lett. 43
  (1979) 1566--1570.
\newblock \href {https://doi.org/10.1103/PhysRevLett.43.1566}
  {\path{doi:10.1103/PhysRevLett.43.1566}}.

\bibitem{CentellesChulia:2018gwr}
S.~Centelles~Chuli\'a, R.~Srivastava, J.~W.~F. Valle, {Seesaw roadmap to
  neutrino mass and dark matter}, Phys. Lett. B 781 (2018) 122--128.
\newblock \href {http://arxiv.org/abs/1802.05722} {\path{arXiv:1802.05722}},
  \href {https://doi.org/10.1016/j.physletb.2018.03.046}
  {\path{doi:10.1016/j.physletb.2018.03.046}}.

\bibitem{Georgi:1974yf}
H.~Georgi, H.~R. Quinn, S.~Weinberg, {Hierarchy of Interactions in Unified
  Gauge Theories}, Phys. Rev. Lett. 33 (1974) 451--454.
\newblock \href {https://doi.org/10.1103/PhysRevLett.33.451}
  {\path{doi:10.1103/PhysRevLett.33.451}}.

\bibitem{Georgi:1974sy}
H.~Georgi, S.~L. Glashow, {Unity of All Elementary Particle Forces}, Phys. Rev.
  Lett. 32 (1974) 438--441.
\newblock \href {https://doi.org/10.1103/PhysRevLett.32.438}
  {\path{doi:10.1103/PhysRevLett.32.438}}.

\bibitem{Weisskopf:1939zz}
V.~F. Weisskopf, {On the Self-Energy and the Electromagnetic Field of the
  Electron}, Phys. Rev. 56 (1939) 72--85.
\newblock \href {https://doi.org/10.1103/PhysRev.56.72}
  {\path{doi:10.1103/PhysRev.56.72}}.

\bibitem{Weinberg:1983xy}
S.~Weinberg, {Charges from Extra Dimensions}, Phys. Lett. B 125 (1983)
  265--269.
\newblock \href {https://doi.org/10.1016/0370-2693(83)91281-9}
  {\path{doi:10.1016/0370-2693(83)91281-9}}.

\bibitem{Djouadi:2013uqa}
A.~Djouadi, L.~Maiani, A.~Polosa, J.~Quevillon, V.~Riquer, {The post-Higgs MSSM
  scenario: Habemus MSSM?}, Eur.Phys.J.C 73 (2013) 2650.
\newblock \href {http://arxiv.org/abs/1307.5205} {\path{arXiv:1307.5205}},
  \href {https://doi.org/10.1140/epjc/s10052-013-2650-0}
  {\path{doi:10.1140/epjc/s10052-013-2650-0}}.

\bibitem{Baglio:2012np}
J.~Baglio, A.~Djouadi, R.~Gr\"ober, M.~M. M\"uhlleitner, J.~Quevillon,
  M.~Spira, {The measurement of the Higgs self-coupling at the LHC: theoretical
  status}, JHEP 04 (2013) 151.
\newblock \href {http://arxiv.org/abs/1212.5581} {\path{arXiv:1212.5581}},
  \href {https://doi.org/10.1007/JHEP04(2013)151}
  {\path{doi:10.1007/JHEP04(2013)151}}.

\bibitem{Weinberg:1978kz}
S.~Weinberg, {Phenomenological Lagrangians}, Physica A 96~(1-2) (1979)
  327--340.
\newblock \href {https://doi.org/10.1016/0378-4371(79)90223-1}
  {\path{doi:10.1016/0378-4371(79)90223-1}}.

\bibitem{deBlas:2017xtg}
J.~de~Blas, J.~C. Criado, M.~Perez-Victoria, J.~Santiago, {Effective
  description of general extensions of the Standard Model: the complete
  tree-level dictionary}, JHEP 03 (2018) 109.
\newblock \href {http://arxiv.org/abs/1711.10391} {\path{arXiv:1711.10391}},
  \href {https://doi.org/10.1007/JHEP03(2018)109}
  {\path{doi:10.1007/JHEP03(2018)109}}.

\bibitem{Weinberg:1987dv}
S.~Weinberg, {Anthropic Bound on the Cosmological Constant}, Phys. Rev. Lett.
  59 (1987) 2607.
\newblock \href {https://doi.org/10.1103/PhysRevLett.59.2607}
  {\path{doi:10.1103/PhysRevLett.59.2607}}.

\end{thebibliography}

\end{document}